\documentclass[aps,showpacs,prb,twocolumn,superscriptaddress]{revtex4-1}

\usepackage{amsmath}
\usepackage[pdftex]{graphicx}
\usepackage{booktabs}
\usepackage{multirow}
\usepackage{hyperref}
\usepackage[mediumspace,mediumqspace,Grey,squaren]{SIunits}
\usepackage{color}

\begin{document}

\newcommand{\cco}{CoCr$_{2}$O$_{4}$}
\newcommand{\cfo}{CoFe$_{2}$O$_{4}$}
\newcommand{\nfo}{NiFe$_{2}$O$_{4}$}
\newcommand{\mao}{MgAl$_{2}$O$_{4}$}
\newcommand{\co}{Co$^{2+}$}
\newcommand{\Cr}{Cr$^{3+}$}
\newcommand{\fe}{Fe$^{3+}$}
\newcommand{\oh}{O$_{h}$}
\newcommand{\td}{T$_{d}$}
\newcommand{\mub}{$\mu_{B}$/f.u.}
\newcommand{\mum}{$\mu_{B}$}
\newcommand{\parallelsum}{\mathbin{\|}}
\newcommand{\NEW}[1]{\color{red} #1 \color{black}}
\newcommand{\OLD}[1]{\color{blue} ( #1) \color{black}}
\newcommand{\AS}[1]{\color{green}  #1 \color{black}}
\newcommand{\CE}[1]{{\bf **CE: #1**}}
\newcommand{\Comm}[1]{\color{red} #1 \color{black}}

\title{Strain-induced magnetic anisotropy in epitaxial thin films of the spinel \cco{} }

\author{J.A. Heuver}
\email{j.a.heuver@rug.nl}
\affiliation{Zernike Institute for Advanced Materials, University of Groningen, Groningen, The Netherlands}
\author{A. Scaramucci}
\email{andrea.scaramucci@psi.ch}
\affiliation{Materials Theory, ETH Z{\"u}rich, Z{\"u}rich, Switzerland}
\affiliation{Laboratory for Scientific Development and Novel Materials, Paul Scherrer Institut, 5235, Villigen PSI, Switzerland}

\author{Y. Blickenstorfer}
\affiliation{Materials Theory, ETH Z{\"u}rich, Z{\"u}rich, Switzerland}
\author{S. Matzen}
\affiliation{Zernike Institute for Advanced Materials, University of Groningen, Groningen, The Netherlands}
\affiliation{ Institut d\'{}Electronique Fondamentale, UMR CNRS, and Universite Paris-Sud, 91405 Orsay, France}

\author{N. A. Spaldin}
\affiliation{Materials Theory, ETH Z{\"u}rich, Z{\"u}rich, Switzerland}

\author{C. Ederer}
\affiliation{Materials Theory, ETH Z{\"u}rich, Z{\"u}rich, Switzerland}

\author{B. Noheda}
\affiliation{Zernike Institute for Advanced Materials, University of Groningen, Groningen, The Netherlands}


\begin{abstract}

We show that the magnetic anisotropy in spinel-structure CoCr$_2$O$_4$ thin films exhibits a strain dependence in which compressive strain induces an out-of-plane magnetic easy axis and tensile strain an in-plane easy axis, exactly opposite to the behavior reported for the related compound CoFe$_2$O$_4$. 
We use density functional theory calculations within the LSDA+U approximation to reproduce and explain the observed behavior. Using second-order perturbation theory, we analyse the anisotropy tensor of the Co$^{2+}$ ions in both octahedral and tetrahedral coordination, allowing us to extend our results to spinels with general arrangements
of Co$^{2+}$ ions. 

\end{abstract}
\maketitle

Thin films with out-of-plane spontaneous magnetization, showing so-called perpendicular magnetic anisotropy, are of great interest for applications such as high density magnetic memories with fast switching~\cite{Mangin2006}. Perpendicular magnetic anisotropy is also needed for spintronic applications, for example in magnetic tunnel junctions, and for low energy current-driven domain wall motion~\cite{Ikeda2010,Yakushiji2010,Fukami2011,Meng2006}.

In a system of finite size, magnetic anisotropy is determined by the balance of magnetocrystalline and shape anisotropy. The former is a bulk property originating from spin-orbit interaction, while the latter originates from magnetic dipole-dipole interaction and depends on the geometry of the sample. 
For the case of thin films, where shape anisotropy always favors in-plane magnetization, it is interesting to understand how the strain affects the magnetocrystalline part. 
Indeed, the way in which epitaxial strain can affect the magnetocrystalline anisotropy has been the subject of intensive investigations~\cite{Dorsey1996,Suzuki1999,Chambers2002,Lisfi2007,Hu2000,Bozorth1955,Gatel2013,Hu2000,Chambers2002,Fritsch2010}. A prominent example of such investigations is the spinel compound \cfo{}, which has a strong magnetostriction coefficient~\cite{Bozorth1955,Suzuki1999,Hu2000,Fritsch2012}. In \cfo{} thin films, it was shown that changing the sign of the strain leads from cooperation to competition of shape and magnetocrystalline anisotropies \cite{Suzuki1999}. 

Compounds with spinel (MgAl$_2$O$_4$-type) structure have chemical formula $AB_2X_4$ where $A$ and $B$ are cations and $X$ represents the ligand anion (usually O, S or Se). In this structure, the cations occupy either the tetrahedrally coordinated\footnote{Here $\mathbf{T}$ and $\mathbf{O}$ do not refer to the Schoenflies notation of the point group but label the type of coordination, respectivelly, tetrahedral and octahedral.} ($\mathbf{T}$) site or the octahedrally
coordinated ($\mathbf{O}$) site. There are twice as many $\mathbf{O}$ sites as $\mathbf{T}$ sites. In a {\it normal} spinel, $A$ and $B$ cations occupy $\mathbf{T}$ and $\mathbf{O}$ sites, respectively, while in an {\it inverse} spinel, half of the $B$ cations occupy the $\mathbf{T}$ sites and the remaining half occupy the $\mathbf{O}$ sites together with the $A$ cations. In general, spinels can be characterised by the degree of inversion (i.e. the concentration of $B$ cations occupying $\mathbf{T}$ sites). 
 
The bulk structure (unstrained case) of the inverse spinel \cfo{}, with a disordered Co$^{2+}$/Fe$^{3+}$ occupancy on the octahedrally coordinated sites, has cubic symmetry (space group $Fd\bar{3}m$). This enforces the quadratic magnetization terms in the  magnetocrystalline anisotropy  to vanish, leaving as lowest order terms the quartic ones, which are typically of smaller size, although they can be appreciably large\cite{Kriegisch2012,Bozorth1955}. However, strain lifts the cubic symmetry, allowing quadratic magnetocrystalline anisotropy, which can either favor in-plane or out-of-plane magnetization. While in \cfo{} compressive strain favors in-plane magnetization, i.e. magnetocrystalline and shape anisotropies cooperate\cite{Hu2000,Suzuki1999,Matzen2011,Gatel2013}, in \cfo{} films grown under tension on MgO, the strain-induced out-of-plane bulk anisotropy is large enough to overcome the shape anisotropy, promoting out-of-plane magnetization\cite{Dorsey1996,Suzuki1999,Chambers2002,Lisfi2007,Fritsch2010}. The magnetocrystalline anisotropy in \cfo{} is expected to be the result of strong spin orbit coupling at the \co{} cation \cite{Slonczewski1958,Tachiki1960} and, thus, a significant effect is expected for different site occupation of the \co{} ions\cite{szotek2006,Fritsch2011}, as observed in \cfo{} under different annealing conditions\cite{Hu2000}.  

Another Co$^{2+}$ compound with cubic spinel structure, \cco{}, raised considerable interest a few years ago as it was discovered to be a multiferroic with strong magnetoelectric coupling when undergoing a transition to a ferrimagnetic conical spiral state\cite{Yamasaki2006,Lawes2006,Tomiyasu2004,Choi2009}. 
As temperature is lowered, \cco{} first shows a transition to a collinear ferrimagnetic state at T$_{N1}$$\sim 95$K, which exists down to T$_{N2}$$\sim 26$K. Below this temperature the ferrimagnetic conical spiral state appears and induces an electric polarization\cite{Choi2009,Yamasaki2006}. At T$\sim14$K an abrupt change in electric polarization occurs\cite{Choi2009,Yamasaki2006}. The fact that it is difficult to grow good quality single crystals leaves some open questions on the detailed magnetic state\cite{Dwight1969,Choi2009}. To date, the epitaxial growth of \cco{} in thin film form has only been reported by a few groups \cite{Luders2004b, Suzuki1996,Hu2002,Liu2014}. No study so far has reported strain engineering of magnetic properties.

\cco{} has a strong tendency to order in a normal spinel structure\cite{Navrotsky1967}, i.e. with
\co{} occupying  $\mathbf{T}$ sites and \Cr{} occupying the $\mathbf{O}$ sites. This is different from the behavior of \cfo{} and \nfo{}, which have an inverse spinel structure, with the \co{} and Ni$^{2+}$ occupying $\mathbf{O}$ sites, while \fe{} cations are distributed in equal proportion among $\mathbf{O}$ and $\mathbf{T}$ sites (although real samples can show partial occupation of \co{} in $\mathbf{T}$ sites due to growth kinetic effects\cite{Hu2000}). Because of the different Co$^{2+}$ distribution in \cco{} and \cfo{}, the effect of strain on magnetic anisotropy is expected to be different in the two compounds. However, it is unknown how these differences will manifest themselves in actual films, since studies on the magnetic anisotropy of strained \cco{}, as well as a comprehensive theoretical treatment for spinels, are lacking. 

In this work, we investigate experimentally and theoretically the strain dependence of magnetocrystalline anisotropy in \cco{}, with the purpose of generalizing the behavior of spinel systems in which Co$^{2+}$ is expected to be the main source of magnetocrystalline anisotropy. After a brief discussion of experimental and computational details, we present our results of magnetic measurements performed for thin films of \cco{} under both compressive and tensile strain. In particular, we find that in \cco{} strain affects magnetocrystalline anisotropy in the opposite way compared to \cfo{}: compressive and tensile strain favor, respectively, out-of-plane and in-plane magnetization.
We then discuss results of {\it ab initio} calculations of the strain dependence of magnetocrystalline anisotropy. Finally, we provide simple arguments, based on second order perturbation theory, which explain the opposite tendencies of the strain dependence of magnetocrystalline anisotropy in \cco{} and \cfo{}.

\section{Experimental and Computational details}

\cco{} thin films were grown using pulsed laser deposition (PLD). The RHEED (Reflection High-Energy Electron Diffraction)-assisted PLD system is equipped with a Lambda Physik COMPex Pro 205 KrF laser with a wavelength of 248 nm. The laser was run with a frequency of 0.5 Hz for 2 hours, using a laser fluence of 3 J/cm$^2$ and a target-substrate distance of 50 mm, which resulted in \cco{} films with thicknesses of about 40 nm. The substrate was  kept at 400-450~$\degree$C during growth. A ceramic \cco{} pellet, sintered by solid state reaction \cite{Mufti2010}, was used as target. The growth took place in a 0.01 mbar oxygen plasma atmosphere, created by an Oxford Scientific MPS-ECR plasma source in order to improve the oxidation of the films\cite{Ma2010}. After deposition, the films were cooled down in 0.5 bar O$_2$ atmosphere with a cooling rate of 5~$\degree$C per minute. The structure of the thin films  was determined by X-ray diffraction (XRD), using an X'pert lab diffractometer. Magnetization was measured using a QD MPMS 7 XL SQUID magnetometer. 
\begin{figure}[t!]
\includegraphics[width=1\columnwidth]{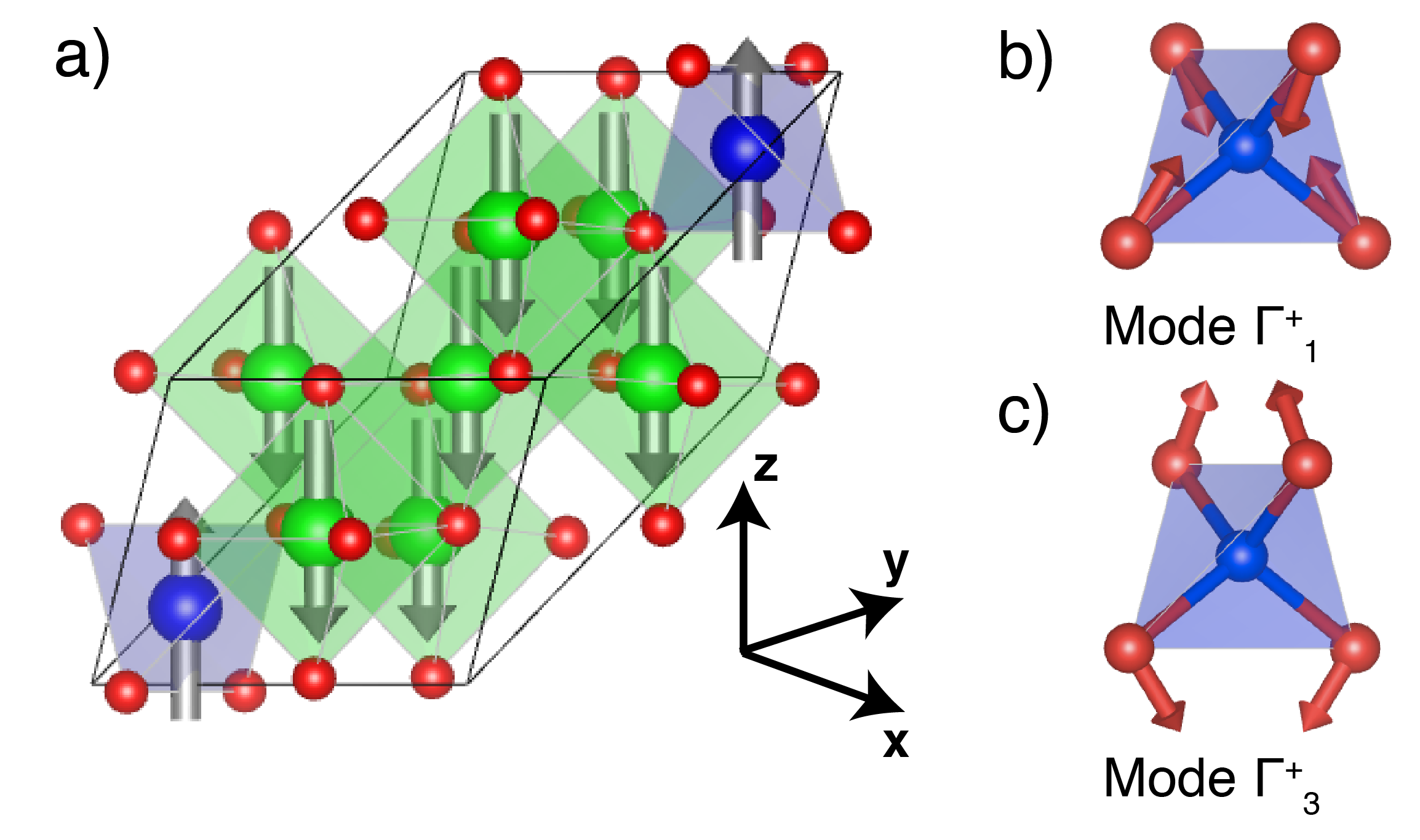}   
\caption{(color online) (a) Crystallographic structure of CoCr$_2$O$_4$. The primitive unit cell of the fcc lattice is shown, containing 2 formula units. Blue, green, and red spheres represent, respectively, Co$^{2+}$, Cr$^{3+}$ and O$^{2-}$ ions. Grey arrows indicate the ferrimagnetic ordering considered in the calculations. Red arrows in panels (b) and (c) show the oxygen displacements for the two modes which appear on relaxing the structure under strain.}
\label{structure}
\end{figure}

Thin films were deposited both on MgO and \mao{} substrates. The substrates were post-annealed at 650~$\degree$C overnight \cite{Eerenstein_thesis} in 1$\times$10$^{-5}$ bar oxygen atmosphere. \mao{} has the spinel crystal structure, like \cco{}, but with a smaller lattice parameter, corresponding to a lattice mismatch\footnote{Here we consider as lattice mismatch $|a_s - a_{CCO}| /a_{CCO}$ where $a_{CCO}$ is the bulk lattice constant of \cco{} and $a_s$ is the  lattice constant of the substrate in the direction relevant for the strain.} of $3.1$ \% that induces a large in-plane compressive strain in the \cco{} films. MgO has the rock-salt structure with a lattice mismatch of $1.1$ \%,  resulting in \cco{} films under tensile strain. 
\begin{figure}[t!]
\includegraphics[width=1\columnwidth]{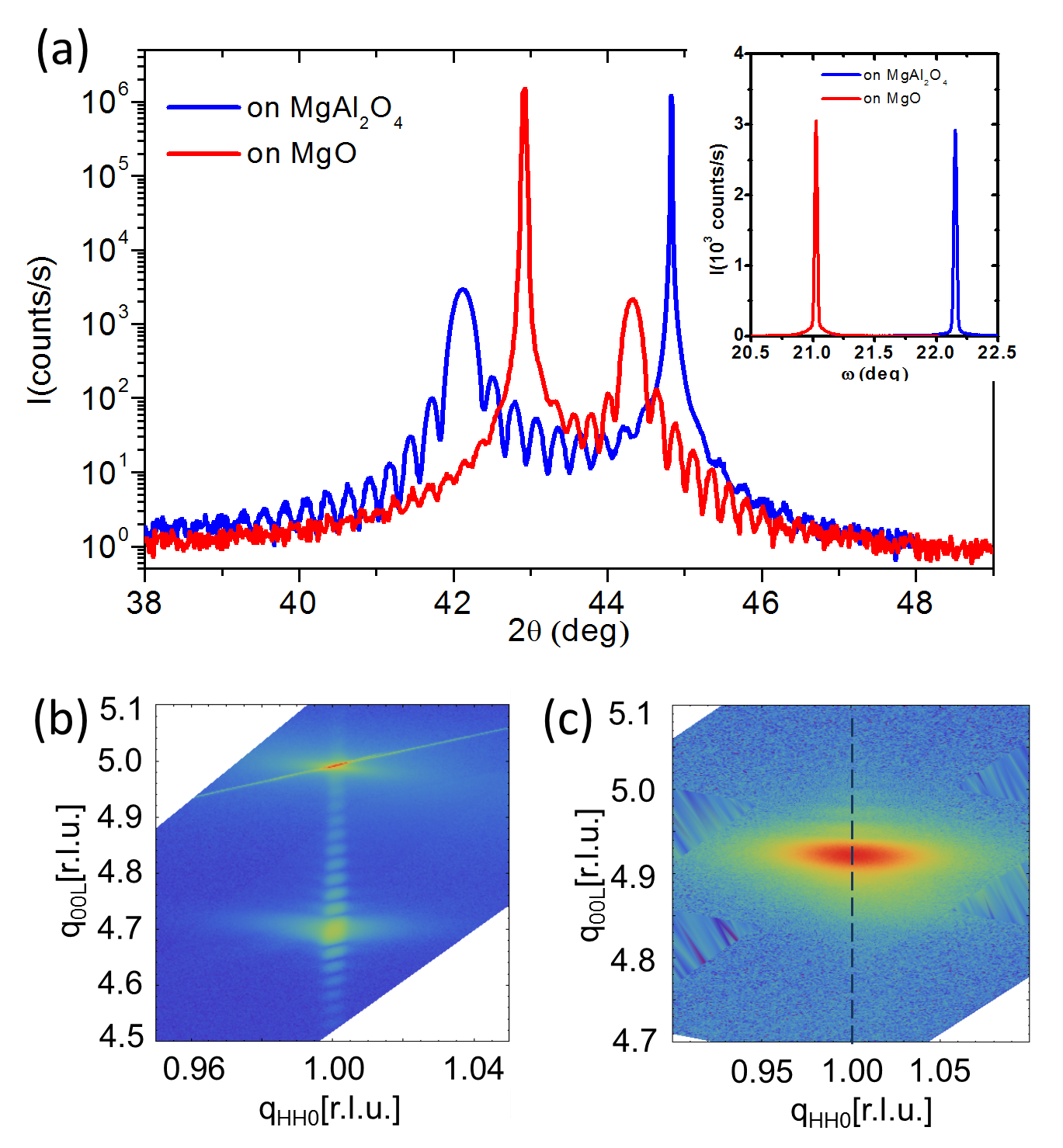}   
\caption{(color online) (a) Specular XRD pattern around the (004) \cco{} reflections. The inset shows the rocking curves around the (004) reflections for both films. XRD Reciprocal space maps (RSMs) around the (115) reflections of the \cco{} films grown on \mao{} (b) and MgO (c). The axes in b) and c) are labelled in reciprocal lattice units (r.l.u.) of the substrate. The vertical dashed line in c) marks the position of the substrate peak (not shown)}.
\label{xrd}
\end{figure}
\\ \indent
{\it Ab initio} calculations were performed using the LSDA+U approach, as implemented in the Vienna Ab initio Simulation Package (VASP), with Projector Augmented  Wave pseudopotentials. 
We use a $7 \times 7 \times 7$ Monkhorst k-point grid and consider the primitive unit cell of the underlying face-centered cubic (fcc) lattice, containing two formula units. 
The values of the effective on-site Coulomb interactions and Hund's couplings for Co$^{2+}$ and Cr$^{3+}$ were set to U$_\text{Co} = 4$ eV, J$^\text{H}_\text{Co} = 1$ eV and U$_\text{Cr} = 3$ eV, J$^\text{H}_\text{Cr} = 1$ eV, respectively. As shown in Ref.~\onlinecite{Ederer2007}, a realistic description of the magnetic properties of \cco{} is obtained using these values. Moreover, for both transition metal ions, the $3s$ and $3p$ states were treated as valence electrons.
The lattice constants and atomic positions, both for the unstrained cubic unit cell and for various values of strain, were obtained by calculating the energy of relaxed structures at different values of the out-of-plane lattice constant and extracting the value corresponding to minimal energy. 
Relaxations were performed considering a collinear ferrimagnetic state in which the spins of the Co$^{2+}$ sublattices are antiparallel to those of the Cr$^{3+}$ sublattices (see Fig.~\ref{structure}a). 
\\ \indent
Calculations to obtain the size of the magnetocrystalline anisotropy were performed using non-collinear spin-polarized calculations, in which the direction of the spins of each magnetic sublattice was constrained and rotated, keeping the ferrimagnetic order described above.

\section{Results and discussion}

\subsection{Experimental results}

Despite the lattice mismatch between both substrates and \cco{}, high quality epitaxial thin films have been grown under both tensile strain, on [001]-oriented MgO substrates, and under compressive strain, on [001]-oriented \mao{} substrates. 
The possibility of applying high amounts of strain to spinel materials was also observed in \cfo{} by Gatel \textit{et al.}\cite{Gatel2013}. The XRD characterization of the films shows a large number of thickness fringes around the film peaks, revealing a superb flatness of the film interfaces (Fig.~\ref{xrd}a). The excellent crystalline quality of the films is shown by the rocking curves of the film peak, displaying full width at half maximum (FWHM) below 0.03$\degree$ (inset of Fig.~\ref{xrd}a). The films on \mao{} and MgO have out-of-plane lattice constants of 8.58 \AA\ and 8.17 \AA, respectively (bulk value is $c = 8.33$\,\AA).

\begin{figure}[t!]
\includegraphics[width=1\columnwidth]{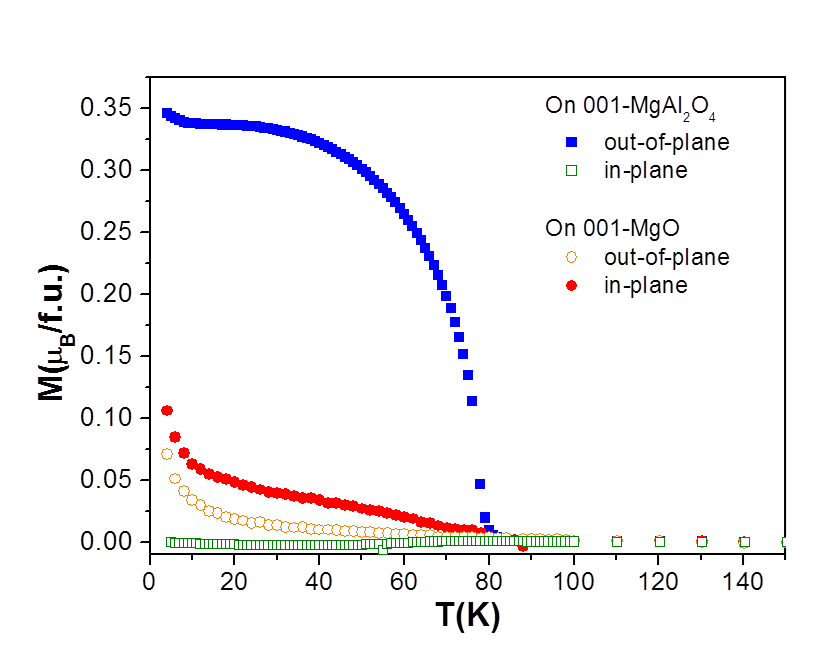}   
\caption{(color online)  Magnetization of strained \cco{} on MgO and \mao{}, measured both in-plane and out-of-plane, in a field of 0.01 T after cooling in 1 T.}
\label{MT}
\end{figure}

The reciprocal space map around the (115)-reflection of \cco{} in Fig.~\ref{xrd}b shows that the in-plane lattice constant of the films are identical to that of the \mao{} substrate, as substrate and film peaks are observed at the same value of q$_{HK0}$. Thus, the films are coherently strained, despite the huge lattice mismatch between the \mao{} and bulk \cco{} of $3.1$\,\%. Despite the relatively low intensity of this reflection, thickness fringes are also visible around the (115)-film peak. 
\\ \indent
According to our specular XRD measurements, \cco{} films grown under tensile strain on MgO display a similar crystal quality to that of the films grown on \mao{}. The (115)-off-specular peak of \cco{} in Fig.~\ref{xrd}c shows that the film on MgO is also fully strained. However, the rocking curve of this peak [scan in the q$_{HH0}$ direction], has a FWHM of 1.2\degree. This is the result of the difference in structure between the rock-salt MgO substrate and the spinel \cco{} film. It is known that the growth of spinel films on rock-salt structure materials can result in the formation of anti-phase boundaries, creating a less homogeneous lattice in the in-plane directions. Antiphase boundaries have, indeed, been observed in our \cco{} thin films using TEM. Their formation and effects will be reported elsewhere. 
\\ \indent
Magnetization of the thin films was measured in both in-plane and out-of-plane geometries. The magnetization of \cco{} is highly anisotropic as shown in Fig.~\ref{MT}. Films on \mao{}, under compressive strain, in a field of 0.01 T applied perpendicular to the film, show the ferrimagnetic transition at 81 K (the bulk value being $T_{N1}\sim 95$\,K). Applying a magnetic field along the [100] in-plane direction induces only a paramagnetic tail, which originates from the substrate contribution, showing that the film has an easy out-of-plane axis. The films on MgO, under tensile strain, show a ferrimagnetic transition at $\sim$80 K under a 0.01 T field applied in the [100] in-plane direction, while no ferrimagnetic transition is observed when the field is applied in the out-of-plane direction. The magnetic anisotropy in \cco{} films, thus,  changes sign depending on the sign of the strain: the magnetic easy axis is out-of-plane in films under compressive strain, while it is in-plane in films under tensile strain. 
As discussed above, the strain-tuning of the magnetic anisotropy has also been observed in films of \cfo{} but in that case the effect is opposite\cite{Suzuki1999}. 

\begin{figure}[t]
\includegraphics[width=1\columnwidth]{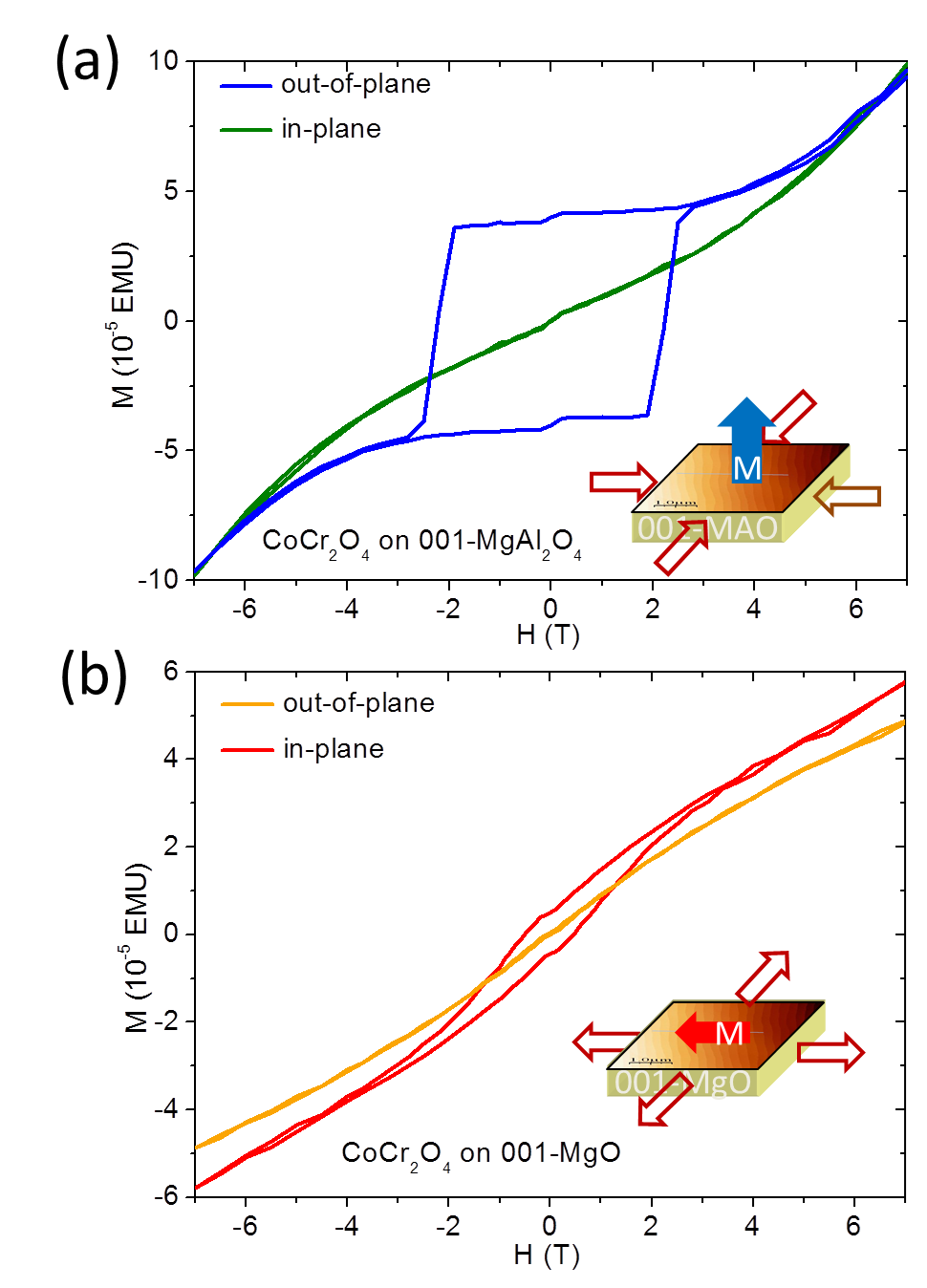}
\caption{(color online)  Magnetic hysteresis loops of \cco{} on \mao{}, under compressive strain (a), and on MgO, under tensile strain (b), at 20 K, measured in the (100) in-plane and (001) out-of-plane directions.}
\label{mh}
\end{figure}

The magnitude of the magnetization is also dependent on the sign of the strain. The low-temperature magnetization of the \cco{} films under compressive strain is 0.34 \mum{} per formula unit (f.u.), while it is only 0.03 \mub{} under tensile strain. 
These values are, respectively, larger and smaller than the 0.08 \mub{} found in single crystals \cite{Yamasaki2006}. Cation inversion is well known to increase or decrease magnetization in spinel structures as for instance in \nfo{} \cite{szotek2006,Luders2005b,Rigato2007}. However, \nfo{} and \cfo{} have a collinear ferrimagnetic structure, whereas in bulk \cco{}, the low temperature magnetic structure is a ferrimagnetic conical spiral in which the low temperature magnetization is not determined by cation inversion (\co{} and \Cr{} both have three unpaired spins) but by the conicities of the magnetic spirals of the various sublattices.

A ferrimagnetic conical spiral can also be present in thin film samples at low temperatures. This, together with a possible dependence of the spiral conicities on the applied strain, might explain the change in the low temperature magnetization.
It is also worth to note that the gyromagnetic tensor is likely to depend weakly on strain, which might contribute to the
discrepancy between the low temperature magnetization of the two samples. Moreover, the magnetization of spinel structures on MgO can be smaller than in bulk as a result of anti-phase boundaries \cite{Margulies1997,Venzke1996,Chambers2002,Lisfi2007}. As mentioned above, anti-phase boundaries were observed in our \cco{} films on MgO, and could thus explain their lower magnetization values. However, since no APB have been observed in our films on \mao{}, it is likely that the increased value of magnetization with respect to bulk \cco{} in this case has other origins, such as changes in conicities.

The strain-dependent magnetic anisotropy is confirmed by the magnetic hysteresis measurements shown in Fig.~\ref{mh}. Films on \mao{} (Fig.~\ref{mh}a) measured with the field along the out-of-plane direction, show a square loop with coercive field of 2.2 T when the diamagnetic slope of the substrate contribution is subtracted, while in the in-plane direction no loop opens. This confirms that the [001] out-of-plane and [100] in-plane axes are, indeed, the easy and hard axes, respectively. The hysteresis loop on MgO has an opening when the field is applied in the [100] in-plane direction while no loop opening is observed in the out-of-plane direction (Fig.~\ref{mh}b). On MgO, the loop in the in-plane direction is not square, which could be a sign of the field not being applied along the easy axis. However, it is well known that films of \cfo{} with anti-phase boundaries have such loop shapes \cite{Margulies1997,Matzen2011}. In our \cco{} films on \mao{}, we observe an extremely large coercive field (unusual for spinels), which most likely originates in the symmetry lowering induced by the strain. Extremely high coercive fields in spinel thin films have also been observed by Yanagihara et al. \cite{Yanagihara2011}.

\subsection{{\it Ab initio} calculation of magnetic anisotropy}

To investigate the microscopic mechanisms leading to the strain dependence of the magnetic anisotropy shown in Fig.~\ref{MT}, we perform {\it ab initio} calculations. We first  calculate the crystal structure under strain by relaxing the out-of-plane lattice constant together with the atomic positions to their lowest energy value for a set of in-plane lattice constants. We then use the relaxed structures to calculate the magnetocrystalline anisotropy. 

In addition to the change of the lattice constants, the structural relaxation of the strained unit cell induces two structural distortions involving oxygen displacements. These correspond to displacement modes of symmetry  $\Gamma^+_1$ and $\Gamma^+_3$.
Mode $\Gamma^+_1$ corresponds to a uniform compression (enlargement) of the oxygen tetrahedra coordinating Co$^{2+}$ when compressive (tensile) strain is applied (see Fig.~\ref{structure}b). 
Mode $\Gamma^+_3$, on the other hand, changes the local symmetry from that of a perfect tetrahedron to that of a tetragonally distorted one, as shown in Fig.~\ref{structure}c). 
We note that the displacements of O$^{2-}$ in mode $\Gamma^+_3$ give rise to shifts in the crystal field splitting similar to those corresponding to the change of the lattice constants when strain is applied. For example, mode $\Gamma^+_3$, when compressive (tensile) strain is applied, shrinks (enlarges) the tetrahedra surrounding Co$^{2+}$  in the $xy$ plane and elongates (shrinks) them along the $z$ axis. 

To estimate magnetocrystalline anisotropies we consider the anisotropic part of the magnetic Hamiltonian
\begin{eqnarray}
H_{A} &= & \sum_{i,\nu} \Delta_{\nu}^{(2)} \left( S_{i,\nu}^z\right)^2 +
\nonumber \\
& &\sum_{i,\nu}\Delta_{\nu}^{(4)}   \left( \left( S_{i,\nu}^x\right)^4 +  \left( S_{i,\nu}^y \right)^4 +\left( S_{i,\nu}^z \right)^4  \right) ,
\label{HAnis}
\end{eqnarray}
where, $i$ and $\nu=1,..,6$ label, respectively, the unit cell and the magnetic sublattices in the unit cell,
$\Delta^{(2)}_{\nu}$  and $\Delta^{(4)}_{\nu}$ are, respectively, the strength of quadratic and quartic anisotropy of spins at sublattice $\nu$ and $S^\alpha$, with $\alpha = x, y, z$, are the spin components along the pseudocubic crystallographic axes. Here we denote with $z$ the direction perpendicular to the film. 
On the one hand, for the bulk case, the first term on the right-hand side of Eq.~\eqref{HAnis} has to vanish due to the cubic symmetry.  On the other hand, while the fourth-order term is always allowed, relatively small tetragonal distortions are expected to make the second-order term dominating. 
\begin{figure}[t]
\includegraphics[width=1\columnwidth]{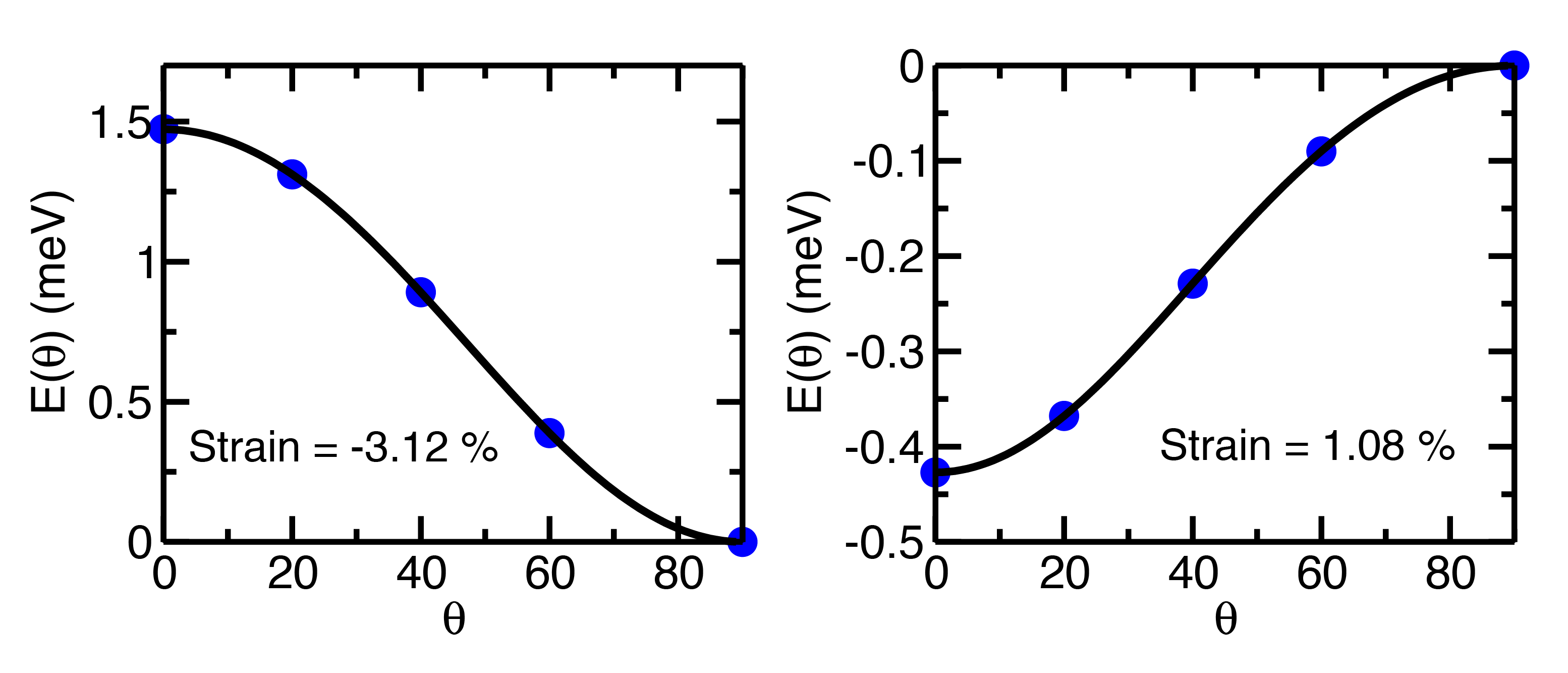}
\caption{(color online) The left and right panels show, respectively, the calculated angle ($\theta$) dependence of the energy (with the energy at $\theta = 90^{\circ}$ subtracted) for $-3.12$\% (compressive) and $1.08$\% (tensile) strain. $\theta = 0$ corresponds to magnetic moment along the $x$ axis (in-plane), while $\theta = 90^{\circ}$ corresponds to magnetic moment along $z$ (out-of-plane). 
The two values of the strain are approximately equal to the experimental ones.}
\label{AnisTwoStrains}
\end{figure}

We map {\it ab initio} energies to the Hamiltonian in Eq.~\eqref{HAnis} using the following procedure (see e.g. Ref.~\onlinecite{Bousquet2012}).
We perform a set of {\it ab initio} calculations where the direction of all the spins is rotated uniformly  (i.e. keeping the magnetic ordering of Fig.~\ref{structure}) from the in-plane direction ($x$) to the out-of-plane direction ($z$) by an angle $\theta$. For these magnetic configurations, Eq.~\eqref{HAnis} yields a $\theta$ dependence of the {\it ab initio} energy per unit cell given by
\begin{equation}  
\epsilon(\theta) = \alpha \cos(\theta)^2 + \beta \left( \cos(\theta)^4 + \sin(\theta)^4\right) + \gamma . 
\label{Etheta}
\end{equation}
Here, $\gamma$ is a constant energy shift which includes all contributions unrelated to magnetocrystalline anisotropy, while $\alpha$ and $\beta$  include, respectively, the quadratic and quartic contributions to spin anisotropy of all six magnetic sublattices in the unit cell. 
\\ \indent
The two panels in Fig.~\ref{AnisTwoStrains} show the numerical results (dots) and the fits (curves) obtained using Eq.\eqref{Etheta} for $\epsilon(\theta)$ for strains close to the experimental values  of $-3.1$ \% (compressive) and $1.1$\% (tensile).
We note that $\epsilon(\theta)$ fits well the {\it ab initio} energies, indicating that the model in Eq.~\eqref{HAnis} describes well the magnetocrystalline anisotropies in this compound. 
Moreover, we find that, in agreement with experimental results, a tensile value of the strain favors a direction of the spins in the film plane (energy minimum at $\theta = 0^{\circ}$),  while compressive strain favors a direction perpendicular to the film plane (energy minimum at $\theta = 90^{\circ}$). 
\begin{figure}[t]
\includegraphics[width=1\columnwidth]{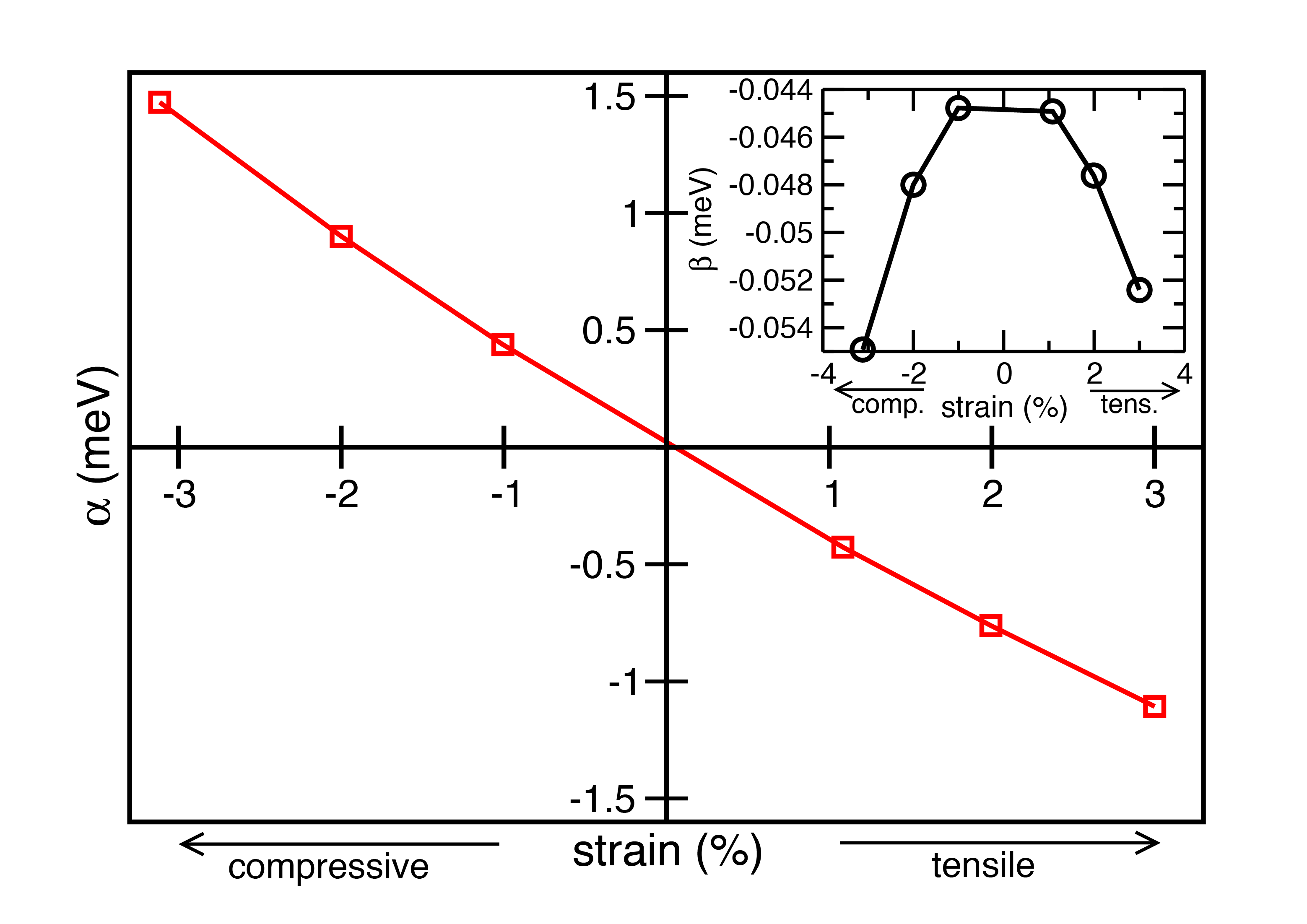}
\caption{(color online) Calculated strain dependence of the constant $\alpha$, as described in the text. The inset shows the strain dependence of the constant $\beta$ related to the quartic anisotropy.}
\label{AlphaVsStrain}
\end{figure}
\\ \indent
For the case of a collinear  ferrimagnetic state, the magnetocrystalline anisotropy part of the microscopic magnetic Hamiltonian, Eq.~\eqref{HAnis}, can be easily related to the  direction of magnetization, i.e. magnetization is favoured along the easy axis anisotropy direction. 
However, for a ferrimagnetic conical spiral state, such straightforward relationship does not always hold. 
In this state, the magnetic ordering of the various sublattices consists of a collinear and a spiral component and the collinear component, which gives rise to the net magnetization, is perpendicular to the plane containing the spiral one. Therefore, whether the collinear or the spiral part of the magnetic ordering is along the easy axis direction might be related to the conicity of the spin ordering, i.e. the relative size of the collinear component with respect to the spiral one. 
Nonetheless, we note that the transition temperatures in Fig.~\ref{MT} are comparable with the transition temperature of the collinear ferrimagnetic state in the bulk ($T_{N1}\sim 95$K). It is thus reasonable to assume that, at least at temperatures below but close to the transition, the magnetic ordering is collinear also for the films and, thus, the magnetization is parallel to the microscopic easy direction given by Eq.\eqref{HAnis}. 
\\ \indent
To generalize the dependence of the magnetocrystalline anisotropy on strain, we extract the values of the constants $\alpha$ and $\beta$ as the strain is tuned. These dependences are shown in Fig.~\ref{AlphaVsStrain}. We note that the strength of the quadratic part ($\alpha$) of the magnetocrystalline anisotropy decreases linearly (to first approximation) as the strain is tuned from compressive to tensile and changes sign for the bulk structure. In contrast, the strength of the quartic term ($\beta$) shows a much weaker dependence on strain.
\\ \indent
It is reasonable to assume that the strongest contribution to the magnetocrystalline anisotropy originates from the Co sublattice due to the $d^7$ electronic configuration of Co$^{2+}$ and the larger value of spin-orbit coupling (which increases with the atomic number) compared to Cr$^{3+}$. 
To verify this assumption, we estimate the separate contributions of the Co$^{2+}$ and Cr$^{3+}$ sublattices by performing the following calculations. 
We constrain the spins of the Cr sublattices to be ferromagnetically aligned  along the $y$ direction and the spins of the Co sublattice to be ferromagnetically aligned along a direction in the $xz$ plane forming an angle $\theta_\text{Co}$ with the $x$ axes. 
As $\theta_\text{Co}$ is changed, the exchange energy within the Co (Cr) sublattices cannot vary,  as all the spins within these sublattices are kept parallel. Moreover, the exchange energy between the Co and Cr sublattices is always zero, as their spins are kept perpendicular to each other. Therefore, the main contribution to the spin part of the Hamiltonian should originate from Eq.~\eqref{HAnis}, which, for such a spin configuration, has the energy dependence  
\begin{equation}
\epsilon(\theta_\text{Co}) = \alpha_{\text{Co}} \cos(\theta_\text{Co})^2 +\beta_{\text{Co}} \left( \cos(\theta_\text{Co})^4 + \sin(\theta_\text{Co})^4 \right).
\end{equation}
Here, $\alpha_{\text{Co}} $ and $\beta_{\text{Co}} $ contain the contribution to the anisotropy of the Co$^{2+}$ sublattice and a constant energy shift is omitted. 
Similar calculations can be performed to obtain the anisotropy contributions of only the Cr$^{3+}$ sublattice, $\alpha_{\text{Cr}}$ and $\beta_{\text{Cr}}$. 
\begin{figure}[t]
\includegraphics[width=1\columnwidth]{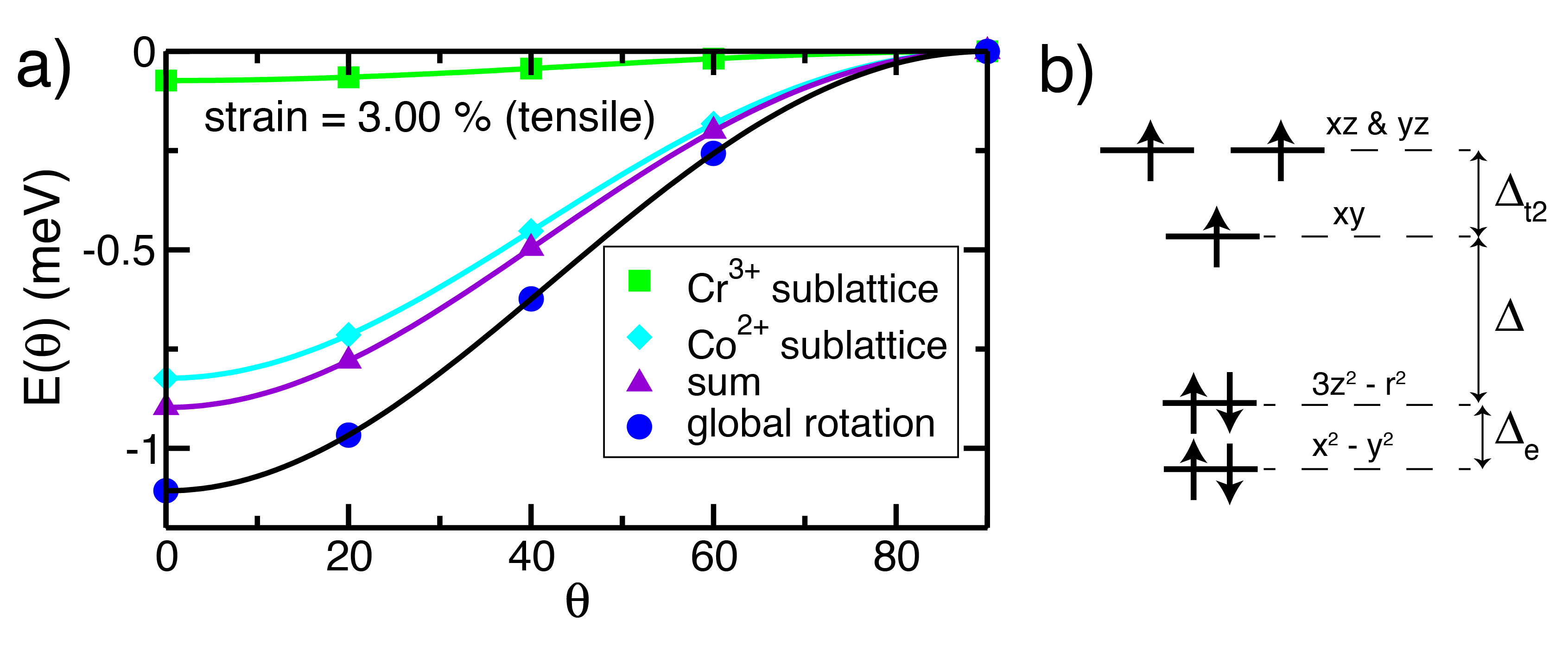}
\caption{(color online) (a) Calculated energy dependence on the angle $\theta$ when the contributions of the Cr$^{3+}$ (green squares) and Co$^{2+}$ (light blue diamonds) are separated as described in the text for 3\% tensile strain. The sum of the two contributions and the energy dependence obtained by a global rotation of all the spins  are shown by (violet) triangles and (blue) dots, respectively. Curves represent the fits obtained as described in the text. (b) Crystal field splitting of a Co$^{2+}$ ion in tetrahedral coordination under compressive strain.}
\label{CoCrSeparated}
\end{figure}
\\ \indent
Figure~\ref{CoCrSeparated}  (a) shows the comparison of the contributions of the Co and Cr sublattices to the magnetocrystalline anisotropy for the case of $3$\% tensile strain. 
In agreement with our initial assumption, the magnetocrystalline anisotropy is dominated by the contribution of Co ions. 
Interestingly, we note that the sum of the contributions of the two sublattices does not give exactly the value obtained for the total magnetocrystalline anisotropy. We further investigated the origin of such difference and excluded possible sizeable contributions from terms such as Dzyaloshinskii-Moriya interaction or anisotropies  of the exchange interactions. A possible source of this discrepancy might arise from a non-local contribution  to the magnetocrystalline anisotropy, e.g. a contribution related to hopping between Co$^{2+}$ and Cr$^{3+}$ which might depend on the relative orientation of the spins of the two sublattices.

\subsection{Second-order perturbation theory for the magnetocrystalline anisotropy}

After establishing that the dominant contribution to magnetocrystalline anisotropy originates from the Co$^{2+}$  sublattices,  we note that the strain dependence of the anisotropy can be understood using a simple picture of isolated Co$^{2+}$ ions coordinated by distorted oxygen tetrahedra.
In this case the Co$^{2+}$ contribution to the first term on the right-hand side of Eq.~\eqref{HAnis} can be obtained by second order perturbation theory in the spin-orbit coupling constant $\lambda$ \cite{Fazekas,Whangbo2008}. 
This gives the expression, 
\begin{equation}
H_{A}= \Delta^{(2)}_{\text{Co}}(S^z)^2=-  \lambda^2 (\Lambda^{zz} - \Lambda^{xx}) (S^z)^2,
\end{equation}
 where we treat the Co spin as a classical vector and 
$\Lambda^{ii} = \sum_{k} \frac{|\langle 0 | L_i | k \rangle |^2 }{\varepsilon_k - \varepsilon_0}$. 
Here, $L_i$ is the $i$-th component of the orbital momentum operator, $| k \rangle$ and $| 0 \rangle$ denote, respectively, the $k$-th excited state and the ground state. To obtain $H_A$, we use the tetragonal symmetry, i.e. $\Lambda^{xx}=\Lambda^{yy}$.\\
We first note that, for the case of a (``normal'') cubic spinel the symmetry of the ions sitting in the $\mathbf{T}$ sites is $\mathbf{T}_d$.
Considering the joint effect of the change of the lattice constants and the mode $\Gamma^+_3$, one obtains (for the case of compressive strain) the splitting of the $d$ levels as sketched in Fig.~\ref{CoCrSeparated} (b). Here, we consider the $d^{7}$ electronic configuration and we assume that  $\Delta > \Delta_{t2} > \Delta_{e} > 0$, as the lobes of $t_2$ states are pointing more directly towards the ligands.
We also note that, in principle, crystal field splittings can be estimated from the {\it ab initio} calculations using Wannier functions. However, it was shown (see e.g. Ref.~\onlinecite{Scaramucci2015}) that this method requires a careful analysis and the so-obtained splittings do not necessarily correspond to a simple point charge model of isolated ions within a crystal field.
By taking into account only the electronic excitations with lowest energy, i.e. only one-electron excitations not violating Hund's rule, we obtain
\begin{equation}
\Delta^{(2)}_{\text{Co-T}} =  \frac{\lambda^2 \left[ - 4 \Delta_{t2} \left(   \Delta +  \Delta_{t2}   \right) +  3 \Delta_{e} \left(   \Delta +  \Delta_{e}   \right)  \right]}{ \left(   \Delta +  \Delta_{t2}   \right)  \left(   \Delta +  \Delta_{e}   \right) \left(   \Delta +  \Delta_{e}   +  \Delta_{t2}\right)  },
\label{analiticAnis}
\end{equation}
where the subscript T indicates that the equation holds for Co$^{2+}$ in tetrahedral coordination.
This expression, for the case of compressive strain, leads to an out-of-plane easy axis anisotropy ($\Delta^{(2)}_{\text{Co}}$ is negative as $\Delta_{t2} > \Delta_{e} $, which favors spin orientation along $z$), in qualitative agreement both with experiments and {\it ab initio} calculations. Moreover, we note that Eq.~\eqref{analiticAnis} holds also for the case of tensile strain, where $\Delta_{t2}$ and $\Delta_{e} $ are expected to be negative with $\Delta_{t2} < \Delta_{e} < 0$. This  gives rise to an easy-plane value of the quadratic anisotropy (positive $\Delta^{(2)}_{\text{Co}}$, which favors spin orientation perpendicular to $z$). 
\\ \indent
Furthermore, we speculate that a similar argument can be used to explain the sign of the magnetostriction in the inverse spinel CoFe$_2$O$_4$. Assuming that also in CoFe$_2$O$_4$ the largest contribution to the magnetocrystalline anisotropy originates from the Co$^{2+}$ ions, we use similar considerations to those used for the tetrahedrally coordinated Co$^{2+}$ to obtain an expression for the anisotropy for the case of octahedral coordination.
It is important to note that in cubic spinel (both normal and inverse) the octahedral sites do not possess symmetry $\mathbf{O}_h$ but are trigonally distorted. Such distortion was previously  shown to play an important role in the surprisingly large quartic magnetic anisotropy found in cubic Co$_x$Fe$_{3-x}$O$_4$\cite{Slonczewski1958,Tachiki1960}.  
Nonetheless, as we are interested in the emergence of second order anisotropy with a strain-induced tetragonal distortion, we consider the case of a perfect $\mathbf{O}_h$ symmetry for the unstrained case. Moreover, we restrict ourselves to the case of  compressive strain where, similarly to the case of CoCr$_2$O$_4$, non-degenerate perturbation theory can be applied\cite{Whangbo2008}.
For compressive strain, we obtain
\begin{equation}
\Delta^{(2)}_{\text{Co-O}} =  \frac{\lambda^2 \left[  \Delta_{t2g} \left(  4  \Delta_{eg}  + 5  \Delta_{t2g}   \right) +
\Delta \left(  \Delta + \Delta_{eg} + 6 \Delta_{t2g}  \right) \right]  }
{  
\Delta_{t2g} \left(   \Delta +  \Delta_{t2g}   \right) \left(   \Delta +  \Delta_{eg}   +  \Delta_{t2g}\right)  
},
\label{analiticAnisFe}
\end{equation}
where the subscript O indicates octahedral coordination, $\Delta_{t2g}= \varepsilon_{xy} -  \varepsilon_{xz/yz}$, $\Delta_{eg}= \varepsilon_{x^2-y^2} -  \varepsilon_{z^2}$, $\Delta= \varepsilon_{z^2} -  \varepsilon_{xy}$ and $\varepsilon_i$ is the on-site energy of the $i$-th $d$ orbital.
For compressive strain of the octahedron, it is reasonable to assume that  $\Delta_{eg}>0$, $\Delta_{t2g}>0$ and $\Delta>0$, which implies that the anisotropy, Eq.~\eqref{analiticAnisFe}, favors in-plane orientation of the spins ($\Delta^{(2)}_{\text{Co-O}}  > 0$).
We stress that in contrast to Eq.~\eqref{analiticAnis}, the expression in Eq.~\eqref{analiticAnisFe} does not hold for tensile strain as the treatment of this case would require the use of degenerate perturbation theory. 

\section{Summary}
We report the growth of fully strained $(001)$ \cco{} films on \mao{} and MgO substrates. Magnetometry experiments reveal that compressively strained \cco{} thin films exhibit a perpendicular magnetization, while films under tensile strain exhibit an in-plane magnetization. The mapping of LSDA+U calculations to the anisotropic part of an effective spin model, Eq.\eqref{HAnis}, shows that (001) compressive strain favours a spin easy axis along the $[001]$ direction while tensile strain favours spin orientation in the film plane. The quadratic anisotropy term changes smoothly as strain is tuned and vanishes in the cubic case. Moreover, these calculations confirm that the main contribution to magnetocrystalline anisotropy is given by the Co$^{2+}$ sublattice. The direction of magnetization is parallel to the easy direction of the microscopic magnetic model for the collinear ferrimagnetic state and for ferrimagnetic conical states with large enough conicity. We note that, for both the samples, the ferrimagnetic state appears at temperature close to T$_{N1}$ below which bulk \cco{} is collinear for a large temperature regime  ($26$K $\lesssim $ T $ \lesssim 95$ K). Therefore, it is reasonable to assume that the magnetization is parallel to the easy direction of model Eq.\eqref{HAnis}. Under this assumption the results obtained with LSDA+U are in agreement with experimental observations.

Furthermore, to explain the behavior of the quadratic part of the magnetocrystalline anisotropy with strain in \cco{}, we give a simple argument based on a perturbative treatment of spin-orbit coupling when the crystal field splitting is modified by strain. Finally, we use these arguments to explain why the strain dependence of anisotropy is opposite in the inverse spinel \cfo{}. The same sign of the strain-induced anisotropy as in \cfo{} has been reported in CoO films \cite{Csiszar2005} (also containing Co$^{2+}$ in octahedral coordination), indicating that the simple argument given in our discussion might not be only specific for spinel structures but can, in principle, be extended to more general cases.

\bibliography{anisotropy}

\begin{thebibliography}{46}%
\makeatletter
\providecommand \@ifxundefined [1]{%
 \@ifx{#1\undefined}
}%
\providecommand \@ifnum [1]{%
 \ifnum #1\expandafter \@firstoftwo
 \else \expandafter \@secondoftwo
 \fi
}%
\providecommand \@ifx [1]{%
 \ifx #1\expandafter \@firstoftwo
 \else \expandafter \@secondoftwo
 \fi
}%
\providecommand \natexlab [1]{#1}%
\providecommand \enquote  [1]{``#1''}%
\providecommand \bibnamefont  [1]{#1}%
\providecommand \bibfnamefont [1]{#1}%
\providecommand \citenamefont [1]{#1}%
\providecommand \href@noop [0]{\@secondoftwo}%
\providecommand \href [0]{\begingroup \@sanitize@url \@href}%
\providecommand \@href[1]{\@@startlink{#1}\@@href}%
\providecommand \@@href[1]{\endgroup#1\@@endlink}%
\providecommand \@sanitize@url [0]{\catcode `\\12\catcode `\$12\catcode
  `\&12\catcode `\#12\catcode `\^12\catcode `\_12\catcode `\%12\relax}%
\providecommand \@@startlink[1]{}%
\providecommand \@@endlink[0]{}%
\providecommand \url  [0]{\begingroup\@sanitize@url \@url }%
\providecommand \@url [1]{\endgroup\@href {#1}{\urlprefix }}%
\providecommand \urlprefix  [0]{URL }%
\providecommand \Eprint [0]{\href }%
\providecommand \doibase [0]{http://dx.doi.org/}%
\providecommand \selectlanguage [0]{\@gobble}%
\providecommand \bibinfo  [0]{\@secondoftwo}%
\providecommand \bibfield  [0]{\@secondoftwo}%
\providecommand \translation [1]{[#1]}%
\providecommand \BibitemOpen [0]{}%
\providecommand \bibitemStop [0]{}%
\providecommand \bibitemNoStop [0]{.\EOS\space}%
\providecommand \EOS [0]{\spacefactor3000\relax}%
\providecommand \BibitemShut  [1]{\csname bibitem#1\endcsname}%
\let\auto@bib@innerbib\@empty
\bibitem [{\citenamefont {Mangin}\ \emph {et~al.}(2006)\citenamefont {Mangin},
  \citenamefont {Ravelosona}, \citenamefont {Katine}, \citenamefont {Carey},
  \citenamefont {Terris},\ and\ \citenamefont {Fullerton}}]{Mangin2006}%
  \BibitemOpen
  \bibfield  {author} {\bibinfo {author} {\bibfnamefont {S.}~\bibnamefont
  {Mangin}}, \bibinfo {author} {\bibfnamefont {D.}~\bibnamefont {Ravelosona}},
  \bibinfo {author} {\bibfnamefont {J.~a.}\ \bibnamefont {Katine}}, \bibinfo
  {author} {\bibfnamefont {M.~J.}\ \bibnamefont {Carey}}, \bibinfo {author}
  {\bibfnamefont {B.~D.}\ \bibnamefont {Terris}}, \ and\ \bibinfo {author}
  {\bibfnamefont {E.~E.}\ \bibnamefont {Fullerton}},\ }\href {\doibase
  10.1038/nmat1595} {\bibfield  {journal} {\bibinfo  {journal} {Nature
  Materials}\ }\textbf {\bibinfo {volume} {5}},\ \bibinfo {pages} {210}
  (\bibinfo {year} {2006})}\BibitemShut {NoStop}%
\bibitem [{\citenamefont {Ikeda}\ \emph {et~al.}(2010)\citenamefont {Ikeda},
  \citenamefont {Miura}, \citenamefont {Yamamoto}, \citenamefont {Mizunuma},
  \citenamefont {Gan}, \citenamefont {Endo}, \citenamefont {Kanai},
  \citenamefont {Hayakawa}, \citenamefont {Matsukura},\ and\ \citenamefont
  {Ohno}}]{Ikeda2010}%
  \BibitemOpen
  \bibfield  {author} {\bibinfo {author} {\bibfnamefont {S.}~\bibnamefont
  {Ikeda}}, \bibinfo {author} {\bibfnamefont {K.}~\bibnamefont {Miura}},
  \bibinfo {author} {\bibfnamefont {H.}~\bibnamefont {Yamamoto}}, \bibinfo
  {author} {\bibfnamefont {K.}~\bibnamefont {Mizunuma}}, \bibinfo {author}
  {\bibfnamefont {H.~D.}\ \bibnamefont {Gan}}, \bibinfo {author} {\bibfnamefont
  {M.}~\bibnamefont {Endo}}, \bibinfo {author} {\bibfnamefont {S.}~\bibnamefont
  {Kanai}}, \bibinfo {author} {\bibfnamefont {J.}~\bibnamefont {Hayakawa}},
  \bibinfo {author} {\bibfnamefont {F.}~\bibnamefont {Matsukura}}, \ and\
  \bibinfo {author} {\bibfnamefont {H.}~\bibnamefont {Ohno}},\ }\href {\doibase
  10.1038/nmat2804} {\bibfield  {journal} {\bibinfo  {journal} {Nature
  Materials}\ }\textbf {\bibinfo {volume} {9}},\ \bibinfo {pages} {721}
  (\bibinfo {year} {2010})}\BibitemShut {NoStop}%
\bibitem [{\citenamefont {Yakushiji}\ \emph {et~al.}(2010)\citenamefont
  {Yakushiji}, \citenamefont {Saruya}, \citenamefont {Kubota}, \citenamefont
  {Fukushima}, \citenamefont {Nagahama}, \citenamefont {Yuasa},\ and\
  \citenamefont {Ando}}]{Yakushiji2010}%
  \BibitemOpen
  \bibfield  {author} {\bibinfo {author} {\bibfnamefont {K.}~\bibnamefont
  {Yakushiji}}, \bibinfo {author} {\bibfnamefont {T.}~\bibnamefont {Saruya}},
  \bibinfo {author} {\bibfnamefont {H.}~\bibnamefont {Kubota}}, \bibinfo
  {author} {\bibfnamefont {A.}~\bibnamefont {Fukushima}}, \bibinfo {author}
  {\bibfnamefont {T.}~\bibnamefont {Nagahama}}, \bibinfo {author}
  {\bibfnamefont {S.}~\bibnamefont {Yuasa}}, \ and\ \bibinfo {author}
  {\bibfnamefont {K.}~\bibnamefont {Ando}},\ }\href {\doibase
  10.1063/1.3524230} {\bibfield  {journal} {\bibinfo  {journal} {Applied
  Physics Letters}\ }\textbf {\bibinfo {volume} {97}},\ \bibinfo {pages}
  {232508} (\bibinfo {year} {2010})}\BibitemShut {NoStop}%
\bibitem [{\citenamefont {Fukami}\ \emph {et~al.}(2011)\citenamefont {Fukami},
  \citenamefont {Suzuki}, \citenamefont {Nakatani}, \citenamefont {Ishiwata},
  \citenamefont {Yamanouchi}, \citenamefont {Ikeda}, \citenamefont {Kasai},\
  and\ \citenamefont {Ohno}}]{Fukami2011}%
  \BibitemOpen
  \bibfield  {author} {\bibinfo {author} {\bibfnamefont {S.}~\bibnamefont
  {Fukami}}, \bibinfo {author} {\bibfnamefont {T.}~\bibnamefont {Suzuki}},
  \bibinfo {author} {\bibfnamefont {Y.}~\bibnamefont {Nakatani}}, \bibinfo
  {author} {\bibfnamefont {N.}~\bibnamefont {Ishiwata}}, \bibinfo {author}
  {\bibfnamefont {M.}~\bibnamefont {Yamanouchi}}, \bibinfo {author}
  {\bibfnamefont {S.}~\bibnamefont {Ikeda}}, \bibinfo {author} {\bibfnamefont
  {N.}~\bibnamefont {Kasai}}, \ and\ \bibinfo {author} {\bibfnamefont
  {H.}~\bibnamefont {Ohno}},\ }\href {\doibase
  http://dx.doi.org/10.1063/1.3558917} {\bibfield  {journal} {\bibinfo
  {journal} {Applied Physics Letters}\ }\textbf {\bibinfo {volume} {98}},\
  \bibinfo {pages} {082504} (\bibinfo {year} {2011})}\BibitemShut {NoStop}%
\bibitem [{\citenamefont {Meng}\ and\ \citenamefont {Wang}(2006)}]{Meng2006}%
  \BibitemOpen
  \bibfield  {author} {\bibinfo {author} {\bibfnamefont {H.}~\bibnamefont
  {Meng}}\ and\ \bibinfo {author} {\bibfnamefont {J.-P.}\ \bibnamefont
  {Wang}},\ }\href {\doibase 10.1063/1.2198797} {\bibfield  {journal} {\bibinfo
   {journal} {Applied Physics Letters}\ }\textbf {\bibinfo {volume} {88}},\
  \bibinfo {pages} {172506} (\bibinfo {year} {2006})}\BibitemShut {NoStop}%
\bibitem [{\citenamefont {Dorsey}\ \emph {et~al.}(1996)\citenamefont {Dorsey},
  \citenamefont {Lubitz}, \citenamefont {Chrisey},\ and\ \citenamefont
  {Horwitz}}]{Dorsey1996}%
  \BibitemOpen
  \bibfield  {author} {\bibinfo {author} {\bibfnamefont {P.~C.}\ \bibnamefont
  {Dorsey}}, \bibinfo {author} {\bibfnamefont {P.}~\bibnamefont {Lubitz}},
  \bibinfo {author} {\bibfnamefont {D.~B.}\ \bibnamefont {Chrisey}}, \ and\
  \bibinfo {author} {\bibfnamefont {J.~S.}\ \bibnamefont {Horwitz}},\ }\href
  {\doibase 10.1063/1.361991} {\bibfield  {journal} {\bibinfo  {journal}
  {Journal of Applied Physics}\ }\textbf {\bibinfo {volume} {79}},\ \bibinfo
  {pages} {6338} (\bibinfo {year} {1996})}\BibitemShut {NoStop}%
\bibitem [{\citenamefont {Suzuki}\ \emph {et~al.}(1999)\citenamefont {Suzuki},
  \citenamefont {Hu}, \citenamefont {van Dover},\ and\ \citenamefont
  {Cava}}]{Suzuki1999}%
  \BibitemOpen
  \bibfield  {author} {\bibinfo {author} {\bibfnamefont {Y.}~\bibnamefont
  {Suzuki}}, \bibinfo {author} {\bibfnamefont {G.}~\bibnamefont {Hu}}, \bibinfo
  {author} {\bibfnamefont {R.~B.}\ \bibnamefont {van Dover}}, \ and\ \bibinfo
  {author} {\bibfnamefont {R.~J.}\ \bibnamefont {Cava}},\ }\href {\doibase
  10.1016/S0304-8853(98)00364-3} {\bibfield  {journal} {\bibinfo  {journal}
  {Journal of Magnetism and Magnetic Materials}\ }\textbf {\bibinfo {volume}
  {191}},\ \bibinfo {pages} {1} (\bibinfo {year} {1999})}\BibitemShut {NoStop}%
\bibitem [{\citenamefont {Chambers}\ \emph {et~al.}(2002)\citenamefont
  {Chambers}, \citenamefont {Farrow}, \citenamefont {Maat}, \citenamefont
  {Toney}, \citenamefont {Folks}, \citenamefont {Catalano}, \citenamefont
  {Trainor},\ and\ \citenamefont {Brown}}]{Chambers2002}%
  \BibitemOpen
  \bibfield  {author} {\bibinfo {author} {\bibfnamefont {S.~A.}\ \bibnamefont
  {Chambers}}, \bibinfo {author} {\bibfnamefont {R.~F.~C.}\ \bibnamefont
  {Farrow}}, \bibinfo {author} {\bibfnamefont {S.}~\bibnamefont {Maat}},
  \bibinfo {author} {\bibfnamefont {M.~F.}\ \bibnamefont {Toney}}, \bibinfo
  {author} {\bibfnamefont {L.}~\bibnamefont {Folks}}, \bibinfo {author}
  {\bibfnamefont {J.~G.}\ \bibnamefont {Catalano}}, \bibinfo {author}
  {\bibfnamefont {T.~P.}\ \bibnamefont {Trainor}}, \ and\ \bibinfo {author}
  {\bibfnamefont {G.~E.}\ \bibnamefont {Brown}},\ }\href {\doibase
  10.1016/S0304-8853(02)00039-2} {\bibfield  {journal} {\bibinfo  {journal}
  {Journal of Magnetism and Magnetic Materials}\ }\textbf {\bibinfo {volume}
  {246}},\ \bibinfo {pages} {124} (\bibinfo {year} {2002})}\BibitemShut
  {NoStop}%
\bibitem [{\citenamefont {Lisfi}\ \emph {et~al.}(2007)\citenamefont {Lisfi},
  \citenamefont {Williams}, \citenamefont {Nguyen}, \citenamefont {Lodder},
  \citenamefont {Coleman}, \citenamefont {Corcoran}, \citenamefont {Johnson},
  \citenamefont {Chang}, \citenamefont {Kumar},\ and\ \citenamefont
  {Morgan}}]{Lisfi2007}%
  \BibitemOpen
  \bibfield  {author} {\bibinfo {author} {\bibfnamefont {A.}~\bibnamefont
  {Lisfi}}, \bibinfo {author} {\bibfnamefont {C.-M.}\ \bibnamefont {Williams}},
  \bibinfo {author} {\bibfnamefont {L.-T.}\ \bibnamefont {Nguyen}}, \bibinfo
  {author} {\bibfnamefont {J.-C.}\ \bibnamefont {Lodder}}, \bibinfo {author}
  {\bibfnamefont {A.}~\bibnamefont {Coleman}}, \bibinfo {author} {\bibfnamefont
  {H.}~\bibnamefont {Corcoran}}, \bibinfo {author} {\bibfnamefont
  {A.}~\bibnamefont {Johnson}}, \bibinfo {author} {\bibfnamefont
  {P.}~\bibnamefont {Chang}}, \bibinfo {author} {\bibfnamefont
  {A.}~\bibnamefont {Kumar}}, \ and\ \bibinfo {author} {\bibfnamefont
  {W.}~\bibnamefont {Morgan}},\ }\href {\doibase 10.1103/PhysRevE.76.054405}
  {\bibfield  {journal} {\bibinfo  {journal} {Physical Review B}\ }\textbf
  {\bibinfo {volume} {76}},\ \bibinfo {pages} {054405} (\bibinfo {year}
  {2007})}\BibitemShut {NoStop}%
\bibitem [{\citenamefont {Hu}\ \emph {et~al.}(2000)\citenamefont {Hu},
  \citenamefont {Choi}, \citenamefont {Eom}, \citenamefont {Harris},\ and\
  \citenamefont {Suzuki}}]{Hu2000}%
  \BibitemOpen
  \bibfield  {author} {\bibinfo {author} {\bibfnamefont {G.}~\bibnamefont
  {Hu}}, \bibinfo {author} {\bibfnamefont {J.-H.}\ \bibnamefont {Choi}},
  \bibinfo {author} {\bibfnamefont {C.-B.}\ \bibnamefont {Eom}}, \bibinfo
  {author} {\bibfnamefont {V.-G.}\ \bibnamefont {Harris}}, \ and\ \bibinfo
  {author} {\bibfnamefont {Y.}~\bibnamefont {Suzuki}},\ }\href {\doibase
  10.1103/PhysRevB.62.R779} {\bibfield  {journal} {\bibinfo  {journal}
  {Physical Review B}\ }\textbf {\bibinfo {volume} {62}},\ \bibinfo {pages}
  {R779} (\bibinfo {year} {2000})}\BibitemShut {NoStop}%
\bibitem [{\citenamefont {Bozorth}\ \emph {et~al.}(1955)\citenamefont
  {Bozorth}, \citenamefont {Tilden},\ and\ \citenamefont
  {Williams}}]{Bozorth1955}%
  \BibitemOpen
  \bibfield  {author} {\bibinfo {author} {\bibfnamefont {R.~M.}\ \bibnamefont
  {Bozorth}}, \bibinfo {author} {\bibfnamefont {E.~F.}\ \bibnamefont {Tilden}},
  \ and\ \bibinfo {author} {\bibfnamefont {A.~J.}\ \bibnamefont {Williams}},\
  }\href {\doibase 10.1103/PhysRev.99.1788} {\bibfield  {journal} {\bibinfo
  {journal} {Physical Review}\ }\textbf {\bibinfo {volume} {99}},\ \bibinfo
  {pages} {1788} (\bibinfo {year} {1955})}\BibitemShut {NoStop}%
\bibitem [{\citenamefont {Gatel}\ \emph {et~al.}(2013)\citenamefont {Gatel},
  \citenamefont {Warot-Fonrose}, \citenamefont {Matzen},\ and\ \citenamefont
  {Moussy}}]{Gatel2013}%
  \BibitemOpen
  \bibfield  {author} {\bibinfo {author} {\bibfnamefont {C.}~\bibnamefont
  {Gatel}}, \bibinfo {author} {\bibfnamefont {B.}~\bibnamefont
  {Warot-Fonrose}}, \bibinfo {author} {\bibfnamefont {S.}~\bibnamefont
  {Matzen}}, \ and\ \bibinfo {author} {\bibfnamefont {J.-B.}\ \bibnamefont
  {Moussy}},\ }\href {\doibase 10.1063/1.4819178} {\bibfield  {journal}
  {\bibinfo  {journal} {Applied Physics Letters}\ }\textbf {\bibinfo {volume}
  {103}},\ \bibinfo {pages} {092405} (\bibinfo {year} {2013})}\BibitemShut
  {NoStop}%
\bibitem [{\citenamefont {Fritsch}\ and\ \citenamefont
  {Ederer}(2010)}]{Fritsch2010}%
  \BibitemOpen
  \bibfield  {author} {\bibinfo {author} {\bibfnamefont {D.}~\bibnamefont
  {Fritsch}}\ and\ \bibinfo {author} {\bibfnamefont {C.}~\bibnamefont
  {Ederer}},\ }\href {\doibase 10.1103/PhysRevB.82.104117} {\bibfield
  {journal} {\bibinfo  {journal} {Physical Review B}\ }\textbf {\bibinfo
  {volume} {82}},\ \bibinfo {pages} {104117} (\bibinfo {year}
  {2010})}\BibitemShut {NoStop}%
\bibitem [{\citenamefont {Fritsch}\ and\ \citenamefont
  {Ederer}(2012)}]{Fritsch2012}%
  \BibitemOpen
  \bibfield  {author} {\bibinfo {author} {\bibfnamefont {D.}~\bibnamefont
  {Fritsch}}\ and\ \bibinfo {author} {\bibfnamefont {C.}~\bibnamefont
  {Ederer}},\ }\href {\doibase 10.1103/PhysRevB.86.014406} {\bibfield
  {journal} {\bibinfo  {journal} {Physical Review B}\ }\textbf {\bibinfo
  {volume} {86}},\ \bibinfo {pages} {014406} (\bibinfo {year}
  {2012})}\BibitemShut {NoStop}%
\bibitem [{Note1()}]{Note1}%
  \BibitemOpen
  \bibinfo {note} {Here $\protect \mathbf {T}$ and $\protect \mathbf {O}$ do
  not refer to the Schoenflies notation of the point group but label the type
  of coordination, respectivelly, tetrahedral and octahedral.}\BibitemShut
  {Stop}%
\bibitem [{\citenamefont {Kriegisch}\ \emph {et~al.}(2012)\citenamefont
  {Kriegisch}, \citenamefont {Ren}, \citenamefont {Sato-Turtelli},
  \citenamefont {M{\"u}ller}, \citenamefont {Gr{\"o}ssinger},\ and\
  \citenamefont {Zhang}}]{Kriegisch2012}%
  \BibitemOpen
  \bibfield  {author} {\bibinfo {author} {\bibfnamefont {M.}~\bibnamefont
  {Kriegisch}}, \bibinfo {author} {\bibfnamefont {W.}~\bibnamefont {Ren}},
  \bibinfo {author} {\bibfnamefont {R.}~\bibnamefont {Sato-Turtelli}}, \bibinfo
  {author} {\bibfnamefont {H.}~\bibnamefont {M{\"u}ller}}, \bibinfo {author}
  {\bibfnamefont {R.}~\bibnamefont {Gr{\"o}ssinger}}, \ and\ \bibinfo {author}
  {\bibfnamefont {Z.}~\bibnamefont {Zhang}},\ }\href@noop {} {\bibfield
  {journal} {\bibinfo  {journal} {Journal of Applied Physics}\ }\textbf
  {\bibinfo {volume} {111}},\ \bibinfo {pages} {07} (\bibinfo {year}
  {2012})}\BibitemShut {NoStop}%
\bibitem [{\citenamefont {Matzen}\ \emph {et~al.}(2011)\citenamefont {Matzen},
  \citenamefont {Moussy}, \citenamefont {Mattana}, \citenamefont {Petroff},
  \citenamefont {Gatel}, \citenamefont {Warot-Fonrose}, \citenamefont {Cezar},
  \citenamefont {Barbier}, \citenamefont {Arrio},\ and\ \citenamefont
  {Sainctavit}}]{Matzen2011}%
  \BibitemOpen
  \bibfield  {author} {\bibinfo {author} {\bibfnamefont {S.}~\bibnamefont
  {Matzen}}, \bibinfo {author} {\bibfnamefont {J.-B.}\ \bibnamefont {Moussy}},
  \bibinfo {author} {\bibfnamefont {R.}~\bibnamefont {Mattana}}, \bibinfo
  {author} {\bibfnamefont {F.}~\bibnamefont {Petroff}}, \bibinfo {author}
  {\bibfnamefont {C.}~\bibnamefont {Gatel}}, \bibinfo {author} {\bibfnamefont
  {B.}~\bibnamefont {Warot-Fonrose}}, \bibinfo {author} {\bibfnamefont {J.~C.}\
  \bibnamefont {Cezar}}, \bibinfo {author} {\bibfnamefont {A.}~\bibnamefont
  {Barbier}}, \bibinfo {author} {\bibfnamefont {M.-a.}\ \bibnamefont {Arrio}},
  \ and\ \bibinfo {author} {\bibfnamefont {P.}~\bibnamefont {Sainctavit}},\
  }\href {\doibase 10.1063/1.3622307} {\bibfield  {journal} {\bibinfo
  {journal} {Applied Physics Letters}\ }\textbf {\bibinfo {volume} {99}},\
  \bibinfo {pages} {052514} (\bibinfo {year} {2011})}\BibitemShut {NoStop}%
\bibitem [{\citenamefont {Slonczewski}(1958)}]{Slonczewski1958}%
  \BibitemOpen
  \bibfield  {author} {\bibinfo {author} {\bibfnamefont {J.~C.}\ \bibnamefont
  {Slonczewski}},\ }\href {\doibase 10.1103/PhysRev.110.1341} {\bibfield
  {journal} {\bibinfo  {journal} {Phys. Rev.}\ }\textbf {\bibinfo {volume}
  {110}},\ \bibinfo {pages} {1341} (\bibinfo {year} {1958})}\BibitemShut
  {NoStop}%
\bibitem [{\citenamefont {Tachiki}(1960)}]{Tachiki1960}%
  \BibitemOpen
  \bibfield  {author} {\bibinfo {author} {\bibfnamefont {M.}~\bibnamefont
  {Tachiki}},\ }\href {\doibase doi:10.1143/PTP.23.1055} {\bibfield  {journal}
  {\bibinfo  {journal} {Prog. Theor. Phys.}\ }\textbf {\bibinfo {volume}
  {23}},\ \bibinfo {pages} {1055} (\bibinfo {year} {1960})}\BibitemShut
  {NoStop}%
\bibitem [{\citenamefont {Szotek}\ \emph {et~al.}(2006)\citenamefont {Szotek},
  \citenamefont {Temmerman}, \citenamefont {K\"{o}dderitzsch}, \citenamefont
  {Svane}, \citenamefont {Petit},\ and\ \citenamefont {Winter}}]{szotek2006}%
  \BibitemOpen
  \bibfield  {author} {\bibinfo {author} {\bibfnamefont {Z.}~\bibnamefont
  {Szotek}}, \bibinfo {author} {\bibfnamefont {W.-M.}\ \bibnamefont
  {Temmerman}}, \bibinfo {author} {\bibfnamefont {D.}~\bibnamefont
  {K\"{o}dderitzsch}}, \bibinfo {author} {\bibfnamefont {A.}~\bibnamefont
  {Svane}}, \bibinfo {author} {\bibfnamefont {L.}~\bibnamefont {Petit}}, \ and\
  \bibinfo {author} {\bibfnamefont {H.}~\bibnamefont {Winter}},\ }\href
  {\doibase 10.1103/PhysRevB.74.174431} {\bibfield  {journal} {\bibinfo
  {journal} {Physical Review B}\ }\textbf {\bibinfo {volume} {74}},\ \bibinfo
  {pages} {174431} (\bibinfo {year} {2006})}\BibitemShut {NoStop}%
\bibitem [{\citenamefont {Fritsch}\ and\ \citenamefont
  {Ederer}(2011)}]{Fritsch2011}%
  \BibitemOpen
  \bibfield  {author} {\bibinfo {author} {\bibfnamefont {D.}~\bibnamefont
  {Fritsch}}\ and\ \bibinfo {author} {\bibfnamefont {C.}~\bibnamefont
  {Ederer}},\ }\href {\doibase 10.1063/1.3631676} {\bibfield  {journal}
  {\bibinfo  {journal} {Applied Physics Letters}\ }\textbf {\bibinfo {volume}
  {99}},\ \bibinfo {pages} {081916} (\bibinfo {year} {2011})}\BibitemShut
  {NoStop}%
\bibitem [{\citenamefont {Yamasaki}\ \emph {et~al.}(2006)\citenamefont
  {Yamasaki}, \citenamefont {Miyasaka}, \citenamefont {Kaneko}, \citenamefont
  {He}, \citenamefont {Arima},\ and\ \citenamefont {Tokura}}]{Yamasaki2006}%
  \BibitemOpen
  \bibfield  {author} {\bibinfo {author} {\bibfnamefont {Y.}~\bibnamefont
  {Yamasaki}}, \bibinfo {author} {\bibfnamefont {S.}~\bibnamefont {Miyasaka}},
  \bibinfo {author} {\bibfnamefont {Y.}~\bibnamefont {Kaneko}}, \bibinfo
  {author} {\bibfnamefont {J.~P.}\ \bibnamefont {He}}, \bibinfo {author}
  {\bibfnamefont {T.}~\bibnamefont {Arima}}, \ and\ \bibinfo {author}
  {\bibfnamefont {Y.}~\bibnamefont {Tokura}},\ }\href@noop {} {\bibfield
  {journal} {\bibinfo  {journal} {Physical Review Letters}\ }\textbf {\bibinfo
  {volume} {96}},\ \bibinfo {pages} {207204} (\bibinfo {year}
  {2006})}\BibitemShut {NoStop}%
\bibitem [{\citenamefont {Lawes}\ \emph {et~al.}(2006)\citenamefont {Lawes},
  \citenamefont {Melot}, \citenamefont {Page}, \citenamefont {Ederer},
  \citenamefont {Hayward}, \citenamefont {Proffen},\ and\ \citenamefont
  {Seshadri}}]{Lawes2006}%
  \BibitemOpen
  \bibfield  {author} {\bibinfo {author} {\bibfnamefont {G.}~\bibnamefont
  {Lawes}}, \bibinfo {author} {\bibfnamefont {B.}~\bibnamefont {Melot}},
  \bibinfo {author} {\bibfnamefont {K.}~\bibnamefont {Page}}, \bibinfo {author}
  {\bibfnamefont {C.}~\bibnamefont {Ederer}}, \bibinfo {author} {\bibfnamefont
  {M.~A.}\ \bibnamefont {Hayward}}, \bibinfo {author} {\bibfnamefont
  {T.}~\bibnamefont {Proffen}}, \ and\ \bibinfo {author} {\bibfnamefont
  {R.}~\bibnamefont {Seshadri}},\ }\href {\doibase 10.1103/PhysRevB.74.024413}
  {\bibfield  {journal} {\bibinfo  {journal} {Phys. Rev. B}\ }\textbf {\bibinfo
  {volume} {74}},\ \bibinfo {pages} {024413} (\bibinfo {year}
  {2006})}\BibitemShut {NoStop}%
\bibitem [{\citenamefont {Tomiyasu}\ \emph {et~al.}(2004)\citenamefont
  {Tomiyasu}, \citenamefont {Fukunaga},\ and\ \citenamefont
  {Suzuki}}]{Tomiyasu2004}%
  \BibitemOpen
  \bibfield  {author} {\bibinfo {author} {\bibfnamefont {K.}~\bibnamefont
  {Tomiyasu}}, \bibinfo {author} {\bibfnamefont {J.}~\bibnamefont {Fukunaga}},
  \ and\ \bibinfo {author} {\bibfnamefont {H.}~\bibnamefont {Suzuki}},\ }\href
  {\doibase 10.1103/PhysRevB.70.214434} {\bibfield  {journal} {\bibinfo
  {journal} {Phys. Rev. B}\ }\textbf {\bibinfo {volume} {70}},\ \bibinfo
  {pages} {214434} (\bibinfo {year} {2004})}\BibitemShut {NoStop}%
\bibitem [{\citenamefont {Choi}\ \emph {et~al.}(2009)\citenamefont {Choi},
  \citenamefont {Okamoto}, \citenamefont {Huang}, \citenamefont {Chao},
  \citenamefont {Lin}, \citenamefont {Chen}, \citenamefont {van Veenendaal},
  \citenamefont {Kaplan},\ and\ \citenamefont {Cheong}}]{Choi2009}%
  \BibitemOpen
  \bibfield  {author} {\bibinfo {author} {\bibfnamefont {Y.-J.}\ \bibnamefont
  {Choi}}, \bibinfo {author} {\bibfnamefont {J.}~\bibnamefont {Okamoto}},
  \bibinfo {author} {\bibfnamefont {D.-J.}\ \bibnamefont {Huang}}, \bibinfo
  {author} {\bibfnamefont {K.-S.}\ \bibnamefont {Chao}}, \bibinfo {author}
  {\bibfnamefont {H.-J.}\ \bibnamefont {Lin}}, \bibinfo {author} {\bibfnamefont
  {C.-T.}\ \bibnamefont {Chen}}, \bibinfo {author} {\bibfnamefont
  {M.}~\bibnamefont {van Veenendaal}}, \bibinfo {author} {\bibfnamefont
  {T.-A.}\ \bibnamefont {Kaplan}}, \ and\ \bibinfo {author} {\bibfnamefont
  {S.-W.}\ \bibnamefont {Cheong}},\ }\href {\doibase
  10.1103/PhysRevLett.102.067601} {\bibfield  {journal} {\bibinfo  {journal}
  {Physical Review Letters}\ }\textbf {\bibinfo {volume} {102}},\ \bibinfo
  {pages} {067601} (\bibinfo {year} {2009})}\BibitemShut {NoStop}%
\bibitem [{\citenamefont {Dwight}(1969)}]{Dwight1969}%
  \BibitemOpen
  \bibfield  {author} {\bibinfo {author} {\bibfnamefont {K.}~\bibnamefont
  {Dwight}},\ }\href {\doibase 10.1063/1.1657574} {\bibfield  {journal}
  {\bibinfo  {journal} {Journal of Applied Physics}\ }\textbf {\bibinfo
  {volume} {40}},\ \bibinfo {pages} {1156} (\bibinfo {year}
  {1969})}\BibitemShut {NoStop}%
\bibitem [{\citenamefont {L\"{u}ders}\ \emph {et~al.}(2004)\citenamefont
  {L\"{u}ders}, \citenamefont {S\'{a}nchez},\ and\ \citenamefont
  {Fontcuberta}}]{Luders2004b}%
  \BibitemOpen
  \bibfield  {author} {\bibinfo {author} {\bibfnamefont {U.}~\bibnamefont
  {L\"{u}ders}}, \bibinfo {author} {\bibfnamefont {F.}~\bibnamefont
  {S\'{a}nchez}}, \ and\ \bibinfo {author} {\bibfnamefont {J.}~\bibnamefont
  {Fontcuberta}},\ }\href {\doibase 10.1016/j.mseb.2003.10.042} {\bibfield
  {journal} {\bibinfo  {journal} {Materials Science and Engineering: B Advanced
  Functional Solid-State Materials}\ }\textbf {\bibinfo {volume} {109}},\
  \bibinfo {pages} {207204} (\bibinfo {year} {2004})}\BibitemShut {NoStop}%
\bibitem [{\citenamefont {Suzuki}\ \emph {et~al.}(1996)\citenamefont {Suzuki},
  \citenamefont {{Van Dover}}, \citenamefont {Gyorgy}, \citenamefont
  {Phillips}, \citenamefont {Korenivski}, \citenamefont {Werder}, \citenamefont
  {Chen}, \citenamefont {Cava}, \citenamefont {Krajewski}, \citenamefont
  {Peck},\ and\ \citenamefont {Do}}]{Suzuki1996}%
  \BibitemOpen
  \bibfield  {author} {\bibinfo {author} {\bibfnamefont {Y.}~\bibnamefont
  {Suzuki}}, \bibinfo {author} {\bibfnamefont {R.~B.}\ \bibnamefont {{Van
  Dover}}}, \bibinfo {author} {\bibfnamefont {E.~M.}\ \bibnamefont {Gyorgy}},
  \bibinfo {author} {\bibfnamefont {J.~M.}\ \bibnamefont {Phillips}}, \bibinfo
  {author} {\bibfnamefont {V.}~\bibnamefont {Korenivski}}, \bibinfo {author}
  {\bibfnamefont {D.~J.}\ \bibnamefont {Werder}}, \bibinfo {author}
  {\bibfnamefont {C.~H.}\ \bibnamefont {Chen}}, \bibinfo {author}
  {\bibfnamefont {R.~J.}\ \bibnamefont {Cava}}, \bibinfo {author}
  {\bibfnamefont {J.~J.}\ \bibnamefont {Krajewski}}, \bibinfo {author}
  {\bibfnamefont {W.~F.}\ \bibnamefont {Peck}}, \ and\ \bibinfo {author}
  {\bibfnamefont {K.~B.}\ \bibnamefont {Do}},\ }\href {\doibase
  10.1063/1.116601} {\bibfield  {journal} {\bibinfo  {journal} {Applied Physics
  Letters}\ }\textbf {\bibinfo {volume} {68}},\ \bibinfo {pages} {714}
  (\bibinfo {year} {1996})}\BibitemShut {NoStop}%
\bibitem [{\citenamefont {Hu}\ and\ \citenamefont {Suzuki}(2002)}]{Hu2002}%
  \BibitemOpen
  \bibfield  {author} {\bibinfo {author} {\bibfnamefont {G.}~\bibnamefont
  {Hu}}\ and\ \bibinfo {author} {\bibfnamefont {Y.}~\bibnamefont {Suzuki}},\
  }\href {\doibase 10.1103/PhysRevLett.89.276601} {\bibfield  {journal}
  {\bibinfo  {journal} {Phys. Rev. Lett.}\ }\textbf {\bibinfo {volume} {89}},\
  \bibinfo {pages} {276601} (\bibinfo {year} {2002})}\BibitemShut {NoStop}%
\bibitem [{\citenamefont {Liu}\ \emph {et~al.}()\citenamefont {Liu},
  \citenamefont {Kareev}, \citenamefont {Cao}, \citenamefont {Liu},
  \citenamefont {Middey}, \citenamefont {Meyers},\ and\ \citenamefont
  {Chakhalian}}]{Liu2014}%
  \BibitemOpen
  \bibfield  {author} {\bibinfo {author} {\bibfnamefont {X.}~\bibnamefont
  {Liu}}, \bibinfo {author} {\bibfnamefont {M.}~\bibnamefont {Kareev}},
  \bibinfo {author} {\bibfnamefont {Y.}~\bibnamefont {Cao}}, \bibinfo {author}
  {\bibfnamefont {J.}~\bibnamefont {Liu}}, \bibinfo {author} {\bibfnamefont
  {S.}~\bibnamefont {Middey}}, \bibinfo {author} {\bibfnamefont
  {D.}~\bibnamefont {Meyers}}, \ and\ \bibinfo {author} {\bibfnamefont
  {J.}~\bibnamefont {Chakhalian}},\ }\href@noop {} {\ }\Eprint
  {http://arxiv.org/abs/1406.0523 [cond-mat.str-el]} {arXiv:1406.0523
  [cond-mat.str-el]} \BibitemShut {NoStop}%
\bibitem [{\citenamefont {Navrotsky}\ and\ \citenamefont
  {Kleppa}(1967)}]{Navrotsky1967}%
  \BibitemOpen
  \bibfield  {author} {\bibinfo {author} {\bibfnamefont {A.}~\bibnamefont
  {Navrotsky}}\ and\ \bibinfo {author} {\bibfnamefont {O.~J.}\ \bibnamefont
  {Kleppa}},\ }\href {\doibase http://dx.doi.org/10.1016/0022-1902(67)80008-3}
  {\bibfield  {journal} {\bibinfo  {journal} {Journal of Inorganic and Nuclear
  Chemistry}\ }\textbf {\bibinfo {volume} {29}},\ \bibinfo {pages} {2701}
  (\bibinfo {year} {1967})}\BibitemShut {NoStop}%
\bibitem [{\citenamefont {Mufti}\ \emph {et~al.}(2010)\citenamefont {Mufti},
  \citenamefont {Nugroho}, \citenamefont {Blake},\ and\ \citenamefont
  {Palstra}}]{Mufti2010}%
  \BibitemOpen
  \bibfield  {author} {\bibinfo {author} {\bibfnamefont {N.}~\bibnamefont
  {Mufti}}, \bibinfo {author} {\bibfnamefont {A.~A.}\ \bibnamefont {Nugroho}},
  \bibinfo {author} {\bibfnamefont {G.~R.}\ \bibnamefont {Blake}}, \ and\
  \bibinfo {author} {\bibfnamefont {T.~T.~M.}\ \bibnamefont {Palstra}},\
  }\href@noop {} {\bibfield  {journal} {\bibinfo  {journal} {Journal of
  Physics: Condensed Matter}\ }\textbf {\bibinfo {volume} {22}},\ \bibinfo
  {pages} {075902} (\bibinfo {year} {2010})}\BibitemShut {NoStop}%
\bibitem [{\citenamefont {Ma}\ \emph {et~al.}(2010)\citenamefont {Ma},
  \citenamefont {Mazumdar}, \citenamefont {Kim}, \citenamefont {Sato},
  \citenamefont {Bao},\ and\ \citenamefont {Gupta}}]{Ma2010}%
  \BibitemOpen
  \bibfield  {author} {\bibinfo {author} {\bibfnamefont {J.~X.}\ \bibnamefont
  {Ma}}, \bibinfo {author} {\bibfnamefont {D.}~\bibnamefont {Mazumdar}},
  \bibinfo {author} {\bibfnamefont {G.}~\bibnamefont {Kim}}, \bibinfo {author}
  {\bibfnamefont {H.}~\bibnamefont {Sato}}, \bibinfo {author} {\bibfnamefont
  {N.~Z.}\ \bibnamefont {Bao}}, \ and\ \bibinfo {author} {\bibfnamefont
  {a.}~\bibnamefont {Gupta}},\ }\href {\doibase 10.1063/1.3488638} {\bibfield
  {journal} {\bibinfo  {journal} {Journal of Applied Physics}\ }\textbf
  {\bibinfo {volume} {108}},\ \bibinfo {pages} {063917} (\bibinfo {year}
  {2010})}\BibitemShut {NoStop}%
\bibitem [{\citenamefont {Eerenstein}()}]{Eerenstein_thesis}%
  \BibitemOpen
  \bibfield  {author} {\bibinfo {author} {\bibfnamefont {W.}~\bibnamefont
  {Eerenstein}},\ }\emph {\bibinfo {title} {Spin-dependent transport across
  anti-phase boundaries in magnetite films}},\ \href@noop {} {Ph.D. thesis},\
  \bibinfo  {school} {University of Groningen}\BibitemShut {NoStop}%
\bibitem [{Note2()}]{Note2}%
  \BibitemOpen
  \bibinfo {note} {Here we consider as lattice mismatch $|a_s - a_{CCO}|
  /a_{CCO}$ where $a_{CCO}$ is the bulk lattice constant of CoCr$_{2}$O$_{4}${}
  and $a_s$ is the lattice constant of the substrate in the direction relevant
  for the strain.}\BibitemShut {Stop}%
\bibitem [{\citenamefont {Ederer}\ and\ \citenamefont
  {Komelj}(2007)}]{Ederer2007}%
  \BibitemOpen
  \bibfield  {author} {\bibinfo {author} {\bibfnamefont {C.}~\bibnamefont
  {Ederer}}\ and\ \bibinfo {author} {\bibfnamefont {M.}~\bibnamefont
  {Komelj}},\ }\href {\doibase 10.1103/PhysRevB.76.064409} {\bibfield
  {journal} {\bibinfo  {journal} {Phys. Rev. B}\ }\textbf {\bibinfo {volume}
  {76}},\ \bibinfo {pages} {064409} (\bibinfo {year} {2007})}\BibitemShut
  {NoStop}%
\bibitem [{\citenamefont {L\"{u}ders}\ \emph {et~al.}(2005)\citenamefont
  {L\"{u}ders}, \citenamefont {Bibes}, \citenamefont {Bobo}, \citenamefont
  {Cantoni}, \citenamefont {Bertacco},\ and\ \citenamefont
  {Fontcuberta}}]{Luders2005b}%
  \BibitemOpen
  \bibfield  {author} {\bibinfo {author} {\bibfnamefont {U.}~\bibnamefont
  {L\"{u}ders}}, \bibinfo {author} {\bibfnamefont {M.}~\bibnamefont {Bibes}},
  \bibinfo {author} {\bibfnamefont {J.~F.}\ \bibnamefont {Bobo}}, \bibinfo
  {author} {\bibfnamefont {M.}~\bibnamefont {Cantoni}}, \bibinfo {author}
  {\bibfnamefont {R.}~\bibnamefont {Bertacco}}, \ and\ \bibinfo {author}
  {\bibfnamefont {J.}~\bibnamefont {Fontcuberta}},\ }\href {\doibase
  10.1103/PhysRevB.71.134419} {\bibfield  {journal} {\bibinfo  {journal}
  {Physical Review B}\ }\textbf {\bibinfo {volume} {71}},\ \bibinfo {pages}
  {134419} (\bibinfo {year} {2005})}\BibitemShut {NoStop}%
\bibitem [{\citenamefont {Rigato}\ \emph {et~al.}(2007)\citenamefont {Rigato},
  \citenamefont {Estrad\'{e}}, \citenamefont {Arbiol}, \citenamefont
  {Peir\'{o}}, \citenamefont {L\"{u}ders}, \citenamefont {Mart\'{\i}},
  \citenamefont {S\'{a}nchez},\ and\ \citenamefont {Fontcuberta}}]{Rigato2007}%
  \BibitemOpen
  \bibfield  {author} {\bibinfo {author} {\bibfnamefont {F.}~\bibnamefont
  {Rigato}}, \bibinfo {author} {\bibfnamefont {S.}~\bibnamefont {Estrad\'{e}}},
  \bibinfo {author} {\bibfnamefont {J.}~\bibnamefont {Arbiol}}, \bibinfo
  {author} {\bibfnamefont {F.}~\bibnamefont {Peir\'{o}}}, \bibinfo {author}
  {\bibfnamefont {U.}~\bibnamefont {L\"{u}ders}}, \bibinfo {author}
  {\bibfnamefont {X.}~\bibnamefont {Mart\'{\i}}}, \bibinfo {author}
  {\bibfnamefont {F.}~\bibnamefont {S\'{a}nchez}}, \ and\ \bibinfo {author}
  {\bibfnamefont {J.}~\bibnamefont {Fontcuberta}},\ }\href {\doibase
  10.1016/j.mseb.2007.07.102} {\bibfield  {journal} {\bibinfo  {journal}
  {Materials Science and Engineering: B}\ }\textbf {\bibinfo {volume} {144}},\
  \bibinfo {pages} {43} (\bibinfo {year} {2007})}\BibitemShut {NoStop}%
\bibitem [{\citenamefont {Margulies}\ \emph {et~al.}(1997)\citenamefont
  {Margulies}, \citenamefont {Parker}, \citenamefont {Rudee}, \citenamefont
  {Spada}, \citenamefont {Chapman}, \citenamefont {Aitchison},\ and\
  \citenamefont {Berkowitz}}]{Margulies1997}%
  \BibitemOpen
  \bibfield  {author} {\bibinfo {author} {\bibfnamefont {D.~T.}\ \bibnamefont
  {Margulies}}, \bibinfo {author} {\bibfnamefont {F.~T.}\ \bibnamefont
  {Parker}}, \bibinfo {author} {\bibfnamefont {M.~L.}\ \bibnamefont {Rudee}},
  \bibinfo {author} {\bibfnamefont {F.~E.}\ \bibnamefont {Spada}}, \bibinfo
  {author} {\bibfnamefont {J.~N.}\ \bibnamefont {Chapman}}, \bibinfo {author}
  {\bibfnamefont {P.~R.}\ \bibnamefont {Aitchison}}, \ and\ \bibinfo {author}
  {\bibfnamefont {A.~E.}\ \bibnamefont {Berkowitz}},\ }\href {\doibase
  10.1103/PhysRevLett.79.5162} {\bibfield  {journal} {\bibinfo  {journal}
  {Physical Review Letters}\ }\textbf {\bibinfo {volume} {79}},\ \bibinfo
  {pages} {5162} (\bibinfo {year} {1997})}\BibitemShut {NoStop}%
\bibitem [{\citenamefont {Venzke}\ \emph {et~al.}(1996)\citenamefont {Venzke},
  \citenamefont {{Van Dover}}, \citenamefont {Phillips}, \citenamefont {Gyory},
  \citenamefont {Siegrist}, \citenamefont {Chen}, \citenamefont {Werder},
  \citenamefont {Fleming}, \citenamefont {Felder}, \citenamefont {Coleman},\
  and\ \citenamefont {Opila}}]{Venzke1996}%
  \BibitemOpen
  \bibfield  {author} {\bibinfo {author} {\bibfnamefont {S.}~\bibnamefont
  {Venzke}}, \bibinfo {author} {\bibfnamefont {R.~B.}\ \bibnamefont {{Van
  Dover}}}, \bibinfo {author} {\bibfnamefont {J.~M.}\ \bibnamefont {Phillips}},
  \bibinfo {author} {\bibfnamefont {E.~M.}\ \bibnamefont {Gyory}}, \bibinfo
  {author} {\bibfnamefont {T.}~\bibnamefont {Siegrist}}, \bibinfo {author}
  {\bibfnamefont {C.~H.}\ \bibnamefont {Chen}}, \bibinfo {author}
  {\bibfnamefont {D.}~\bibnamefont {Werder}}, \bibinfo {author} {\bibfnamefont
  {R.~M.}\ \bibnamefont {Fleming}}, \bibinfo {author} {\bibfnamefont {R.~J.}\
  \bibnamefont {Felder}}, \bibinfo {author} {\bibfnamefont {E.}~\bibnamefont
  {Coleman}}, \ and\ \bibinfo {author} {\bibfnamefont {R.}~\bibnamefont
  {Opila}},\ }\href {\doibase 10.1557/JMR.1996.0153} {\bibfield  {journal}
  {\bibinfo  {journal} {Journal of Materials Research}\ }\textbf {\bibinfo
  {volume} {11}},\ \bibinfo {pages} {1187} (\bibinfo {year}
  {1996})}\BibitemShut {NoStop}%
\bibitem [{\citenamefont {Yanagihara}\ \emph {et~al.}(2011)\citenamefont
  {Yanagihara}, \citenamefont {Uwabo}, \citenamefont {Minagawa}, \citenamefont
  {Kita},\ and\ \citenamefont {Hirota}}]{Yanagihara2011}%
  \BibitemOpen
  \bibfield  {author} {\bibinfo {author} {\bibfnamefont {H.}~\bibnamefont
  {Yanagihara}}, \bibinfo {author} {\bibfnamefont {K.}~\bibnamefont {Uwabo}},
  \bibinfo {author} {\bibfnamefont {M.}~\bibnamefont {Minagawa}}, \bibinfo
  {author} {\bibfnamefont {E.}~\bibnamefont {Kita}}, \ and\ \bibinfo {author}
  {\bibfnamefont {N.}~\bibnamefont {Hirota}},\ }\href {\doibase
  10.1063/1.3566079} {\bibfield  {journal} {\bibinfo  {journal} {Journal of
  Applied Physics}\ }\textbf {\bibinfo {volume} {109}},\ \bibinfo {pages}
  {07C122} (\bibinfo {year} {2011})}\BibitemShut {NoStop}%
\bibitem [{\citenamefont {Weingart}\ \emph {et~al.}(2012)\citenamefont
  {Weingart}, \citenamefont {Spaldin},\ and\ \citenamefont
  {Bousquet}}]{Bousquet2012}%
  \BibitemOpen
  \bibfield  {author} {\bibinfo {author} {\bibfnamefont {C.}~\bibnamefont
  {Weingart}}, \bibinfo {author} {\bibfnamefont {N.}~\bibnamefont {Spaldin}}, \
  and\ \bibinfo {author} {\bibfnamefont {E.}~\bibnamefont {Bousquet}},\ }\href
  {\doibase 10.1103/PhysRevB.86.094413} {\bibfield  {journal} {\bibinfo
  {journal} {Phys. Rev. B}\ }\textbf {\bibinfo {volume} {86}},\ \bibinfo
  {pages} {094413} (\bibinfo {year} {2012})}\BibitemShut {NoStop}%
\bibitem [{\citenamefont {Fazekas}(1999)}]{Fazekas}%
  \BibitemOpen
  \bibfield  {author} {\bibinfo {author} {\bibfnamefont {P.}~\bibnamefont
  {Fazekas}},\ }\href@noop {} {\emph {\bibinfo {title} {Lecture notes on
  electron correlation and magnetism}}}\ (\bibinfo  {publisher} {World
  Scientific},\ \bibinfo {year} {1999})\BibitemShut {NoStop}%
\bibitem [{\citenamefont {Dai}\ \emph {et~al.}(2008)\citenamefont {Dai},
  \citenamefont {Xiang},\ and\ \citenamefont {Whangbo}}]{Whangbo2008}%
  \BibitemOpen
  \bibfield  {author} {\bibinfo {author} {\bibfnamefont {D.}~\bibnamefont
  {Dai}}, \bibinfo {author} {\bibfnamefont {H.}~\bibnamefont {Xiang}}, \ and\
  \bibinfo {author} {\bibfnamefont {M.-H.}\ \bibnamefont {Whangbo}},\ }\href
  {\doibase 10.1002/jcc.21011} {\bibfield  {journal} {\bibinfo  {journal}
  {Journal of Computational Chemistry}\ }\textbf {\bibinfo {volume} {29}},\
  \bibinfo {pages} {2187} (\bibinfo {year} {2008})}\BibitemShut {NoStop}%
\bibitem [{\citenamefont {Scaramucci}\ \emph {et~al.}(2015)\citenamefont
  {Scaramucci}, \citenamefont {Ammann}, \citenamefont {Spaldin},\ and\
  \citenamefont {Ederer}}]{Scaramucci2015}%
  \BibitemOpen
  \bibfield  {author} {\bibinfo {author} {\bibfnamefont {A.}~\bibnamefont
  {Scaramucci}}, \bibinfo {author} {\bibfnamefont {J.}~\bibnamefont {Ammann}},
  \bibinfo {author} {\bibfnamefont {N.~A.}\ \bibnamefont {Spaldin}}, \ and\
  \bibinfo {author} {\bibfnamefont {C.}~\bibnamefont {Ederer}},\ }\href
  {http://stacks.iop.org/0953-8984/27/i=17/a=175503} {\bibfield  {journal}
  {\bibinfo  {journal} {Journal of Physics: Condensed Matter}\ }\textbf
  {\bibinfo {volume} {27}},\ \bibinfo {pages} {175503} (\bibinfo {year}
  {2015})}\BibitemShut {NoStop}%
\bibitem [{\citenamefont {Csiszar}\ \emph {et~al.}(2005)\citenamefont
  {Csiszar}, \citenamefont {Haverkort}, \citenamefont {Hu}, \citenamefont
  {Tanaka}, \citenamefont {Hsieh}, \citenamefont {Lin}, \citenamefont {Chen},
  \citenamefont {Hibma},\ and\ \citenamefont {Tjeng}}]{Csiszar2005}%
  \BibitemOpen
  \bibfield  {author} {\bibinfo {author} {\bibfnamefont {S.-I.}\ \bibnamefont
  {Csiszar}}, \bibinfo {author} {\bibfnamefont {M.-W.}\ \bibnamefont
  {Haverkort}}, \bibinfo {author} {\bibfnamefont {Z.}~\bibnamefont {Hu}},
  \bibinfo {author} {\bibfnamefont {A.}~\bibnamefont {Tanaka}}, \bibinfo
  {author} {\bibfnamefont {H.-H.}\ \bibnamefont {Hsieh}}, \bibinfo {author}
  {\bibfnamefont {H.-J.}\ \bibnamefont {Lin}}, \bibinfo {author} {\bibfnamefont
  {C.-T.}\ \bibnamefont {Chen}}, \bibinfo {author} {\bibfnamefont
  {T.}~\bibnamefont {Hibma}}, \ and\ \bibinfo {author} {\bibfnamefont {L.-H.}\
  \bibnamefont {Tjeng}},\ }\href {\doibase 10.1103/PhysRevLett.95.187205}
  {\bibfield  {journal} {\bibinfo  {journal} {Physical Review Letters}\
  }\textbf {\bibinfo {volume} {95}},\ \bibinfo {pages} {187205} (\bibinfo
  {year} {2005})}\BibitemShut {NoStop}%
\end{thebibliography}%
\end{document}